\documentclass[12pt, draftclsnofoot, onecolumn]{IEEEtran}

\usepackage[utf8]{inputenc}
\usepackage{amsmath,amssymb,amsfonts}
\usepackage{algorithm}
\usepackage{algorithmic}
\usepackage{graphicx}
\usepackage{textcomp}
\usepackage{psfrag}
\usepackage{dsfont}
\usepackage{subfig}
\usepackage{multirow}
\DeclareMathOperator*{\argmax}{arg \: max} 
\DeclareMathOperator*{\argmin}{arg \: min} 
\usepackage{cite}

\begin{document}

\title{
On hybrid precoder/combiner for downlink mmWave massive MU-MIMO systems 
}

\author{Alvaro Javier Ortega, Raimundo Sampaio-Neto and Rodrigo Pereira David

\thanks{Manuscript received XXX, XX, 2018; revised XXX, XX, 2018. This work was supported by the Brazilian National Council for Scientific and Technological Development (CNPq). The review of this paper was coordinated by XXXX.}
\thanks{A. J. Ortega and R. Sampaio-Neto are with the Center for Studies in Telecommunications (CETUC), Pontifical Catholic University of Rio de Janeiro (PUC-Rio), 22451-900 Rio de Janeiro, Brazil (e-mail: javier.ortega@cetuc.puc-rio.br; raimundo@cetuc.puc-rio.br).}
\thanks{R. D. Pereira is with the Laboratory of Telecommunications in the Brazilian Institute of metrology (INMETRO), 25250-020 Rio de Janeiro, Brazil (e-mail: rpdavid@inmetro.gov.br).}}
\markboth{}
{}


\maketitle

\begin{abstract}
We propose four hybrid combiner/precoder for downlink mmWave massive MU-MIMO systems. The design of a hybrid combiner/precoder is divided in two parts, analog and digital. The system baseband model shows that the  signal processed by the mobile station can be interpreted as a received signal in the presence of colored Gaussian noise, therefore, since the digital part of the combiner and precoder do not have constraints for their generation, their designs can be based on any traditional signal processing that takes into account this kind of noise. To the best of our knowledge, this was not considered by previous works. A more realistic and appropriate design is described in this paper. Also, the  approaches adopted in the literature for the designing of the combiner'/precoder' analog parts do not try to avoid or even reduce the inter user/symbol interference, they concentrate on increasing the signal-to-noise ratio (SNR). We propose a simple solution that decreases the interference while maintaining large SNR. In addition, one of the proposed hybrid combiners reaches the maximum value of our objective function according with the Hadamard's inequality. Numerical results illustrate the BER performance improvements resulting from our proposals. In addition, a simple detection approach can be used for data estimation without significant performance loss.
\end{abstract}


\IEEEpeerreviewmaketitle

\section{Introduction}

Wireless communications systems exploiting millimeter wave (mmWave) frequencies are thought to be a core technology that will enable the deployment of the fifth generation (5G) cellular system. Broadband mmWave systems promise significant increase in the data rates due to the extremely wide bandwidths available in the mmWave spectrum. The adverse channel conditions at mmWave frequencies make the communication a hard challenge, this can however be compensate by means of using a large number of antennas that results in large array gain \cite{1,8}. Nevertheless, the conventional fully digital precoding techniques require a dedicated radio frequency (RF) chain for each antenna element, which is impractical at present due to the high cost and power consumption. Hence, it is desirable to design economical hardware that will utilize the potential gain from a large number of cheap antenna elements using a small number of expensive RF chains. Thus, processing schemes with less RF chains than antenna elements a.k.a. hybrid processing have been proposed \cite{29}. 

Several approaches have been considered for hybrid processing. The work in \cite{1} is one of the most popular in the literature, the authors exploit the spatial structure of mmWave channels to formulate the precoding/combining problem as a sparse reconstruction problem and they proposed the principle of basis pursuit as tool for its solution. This idea motivated other authors to continue developing hybrid precoders based on sparse reconstruction, e.g., \cite{3_12_11,1just,2just}. Similar approaches were taken in \cite{3_14, n11_8, n11_9}, where a  digital precoder/combiner is defined and then a complex algorithm is proposed to find a hybrid approximation, e.g., gradient descent, considering a weighted sum mean square error (WSMSE) minimization problem or using the orthogonal matching pursuit algorithm. In \cite{n11_10}, a algorithm based on manifold optimization is proposed. In each iteration of the algorithm, it assumes a given digital precoder and develops a conjugate gradient method to find an analog precoder that is a local minimizer of the approximation gap from the fully-digital one. Next, the digital precoder is computed using a least squares solution. This method achieves good performance but suffers from high complexity and run time. More recently works as \cite{21,29, m3} lead to more successful methodologies, which are described briefly in this paper. 

We propose four hybrid combiner/precoder for downlink mmWave massive MU-MIMO systems. The designing of a hybrid combiner/precoder is divided in two parts, analog and digital. The system model in baseband shows that the signal processed by the mobile station can be interpreted as a received signal in the presence of colored Gaussian noise, therefore, since the digital part of the combiner in the receiver and in the precoder do not have constraints for their generation, they can be designed using any traditional signal processing approaches that take into account this kind of noise. However, to the best of our knowledge, previous works do not consider colored noise, e.g., \cite{13,21,29}. A more realistic and appropriate design is described in this paper. Two of our proposals consist in the improvement of the digital part of the works in \cite{21,m3} using that stated before.   
 
On the other hand, typical approaches for the designing of the analog parts of both the combiner and the precoder are focused on increasing the detection signal-to-noise ratio (SNR) without trying to reduce the inter user/symbol interference. In this paper we propose two simple solutions that are able to decrease the inter symbol interference while keeping SNR large. The first proposal consists in the improvement of the iterative algorithm proposed in \cite{29} through a recursive algorithm. The second is based on single value decomposition (SVD), and reaches excellent performance with much less complexity. In addition one of the the proposed hybrid combiners reaches the maximum value of our objective function according with the Hadamard's inequality. Numerical results in different environments show the improvement obtained through our proposals in relation to the considered hybrid combiner/precoder \cite{21,29, m3}.

The remaining of this paper is organized as follows: Section \ref{Sec.systemmodel} and Section \ref{Sec.systemchannel} describe the system model and channel model, respectively;  Section \ref{sec.HBapproaches} resumes the hybrid design approaches described in \cite{21,29,m3}; Section \ref{sec.pro1}, \ref{sec.pro2.3} and \ref{sec.pro4} are dedicated to  describe our proposals. Section \ref{sec.detection} presents four sub-optimal data detection approaches. In Section \ref{Sec.numerical} simulations results are shown; and finally, in Section \ref{Sec.conclusions} some conclusions wrap up this paper.

The following notation is used throughout the paper: $\mathbb{C}$ denotes the field of complex numbers; $\mathcal{A}$ is a set; $\mathbf{A}$ is a matrix; $\mathbf{a}$ is a vector; $a$ is a scalar; $\mathbf{A}_{a,b}$,   $\mathbf{A}_{a,:}$,  $\mathbf{A}_{:,b}$,  denote the $(a,b)$-th entry, $a$-th row, and $b$-th column of the matrix $\mathbf{A}$, respectively; $\mathbf{1}_{a,b}$ is the $a$x$b$ all ones matrix; $\mathbf{I}_N$ is the $N$x$N$ identity matrix; $\mathrm{tr}\{\mathbf{A}\}$ returns the trace of matrix $\mathbf{A}$; $\parallel . \parallel_p $ is the $p$-norm, for the euclidean norm case, $p=2$, the under-index is avoided; $\mathrm{det}(.)$ represents the determinant function; $\mid \mathbf{A} \mid $ returns the product of the nonzero eigenvalues of the square matrix $\mathbf{A} $; $\otimes$ is the Kronecker product; $(.)^T$ and $(.)^H$ denote the transpose and conjugate transpose, respectively; $\mathbb{E}[.]$ is the expectation operator; $\mathcal{CN}(m,\sigma^2)$ denotes a complex Gaussian random variable with mean $m$ and variance $\sigma^2$; and the function $\mathbf{\Psi(\mathbf{A})} $ returns the entries of the matrix $ \mathbf{A} \in \mathbb{C} ^{n \times m} $ normalized to magnitude 1, i.e.,  $(\mathbf{\Psi(\mathbf{A})})_{i,j} = \frac{\mathbf{A}_{i,j}}{ \parallel \mathbf{A}_{i,j} \parallel } , i=1,...,n , j=1,....m  $

\section{System model}
\label{Sec.systemmodel}
We consider downlink mmWave MU-MIMO systems using HB in the base station (BS) and in each mobile station (MS).  The HB in the BS can be represented by the product between the RF beamformer, $\mathbf{F}_{RF} \in \mathbb{C}^{N_t \times N_{RF_t}}$, and the baseband beamformer, $\mathbf{F}_{BB} \in \mathbb{C}^{N_{RF_t} \times KN_s}$. There are $K$ users equipped with $N_r$ antennas and $N_{RF_r}$ RF chains to process $N_s$ streams. The BS has $N_t$ antennas and sends $KN_s$ streams simultaneously using $N_{RF_t}$ RF chains, where $N_{RF_t}$ satisfies $KN_s \leq N_{RF_t} \leq N_t$. If $ N_{RF_t}$ is equal to $N_t$, the BS performs digital beamformer \cite{13}.

Power normalization is satisfied such that $\parallel \mathbf{F}_{RF} \mathbf{F}_{BB} \parallel^2_F = KN_s$. Then the received signal by the user $k$, $\mathbf{r}_k \in \mathbb{C}^{N_r \times 1}$, is expressed as

\begin{equation}
\mathbf{r}_k = \mathbf{H}_k \mathbf{F}_{RF} \mathbf{F}_{BB}\mathbf{s} + \mathbf{n}_k
\end{equation}

\noindent where $\mathbf{H}_k \in \mathbb{C}^{N_r \times N_t}$ denotes the channel matrix from the BS to the user $k$ satisfying $\mathbb{E} [\parallel \mathbf{H}_k \parallel_F^2 ] = N_t N_r$; $\mathbf{n}_k \in \mathbb{C}^{Nr \times 1}$ is a complex Gaussian noise vector with zero-mean and covariance matrix $\sigma_n^2 \mathbf{I}_{N_r}$, i.e., $\mathcal{CN}(0,\sigma^2_n\mathbf{I}_{N_r})$; $\mathbf{s} \in \mathbb{Q}^{KN_s \times 1}$ is the data stream vector expressed as the concatenation of the user's stream vectors such that $\mathbf{s} = \left[\mathbf{s}_1^T, \mathbf{s}_2^T, ..., \mathbf{s}_K^T \right]^T$ with $\mathbb{E}[ \mathbf{ss}^H ] = \mathbf{I}_{KN_s}$ and whose entries belong to a constellation $\mathbb{Q}$. The analog part of the precoder, $\mathbf{F}_{RF}$, is implemented by phase shifters, satisfying $ \parallel \left( \mathbf{F}_{RF} \right)_{i,j} \parallel = \frac{1}{\sqrt{N_t}}$.

The receiver uses its $N_{RF_r}$ RF chains and analog phase shifters to obtain the processed received signal

\begin{equation}
\label{eq.prs}
\mathbf{y}_k = \mathbf{W}_{BB_k}^H  \mathbf{W}_{RF_k}^H   \mathbf{H}_k \mathbf{F}_{RF}\mathbf{F}_{BB} \mathbf{s} + \mathbf{W}_{BB_k}^H  \mathbf{W}_{RF_k}^H   \mathbf{n}_k
\end{equation}

\noindent where $\mathbf{W}_{RF_k} \in \mathbb{C}^{N_r \times N_{RF_r}}$ is the RF combining matrix and $\mathbf{W}_{BB_k} \in \mathbb{C}^{N_{RF_r} \times N_s}$ denotes the baseband combining matrix of the user $k$. Similarly to the RF precoder, $\mathbf{W}_{RF_k}$ is implemented using phase shifters and therefore $ \parallel \left( \mathbf{W}_{RF_k} \right)_{i,j} \parallel = \frac{1}{\sqrt{N_r}}$\cite{1}.

Equation (\ref{eq.prs}) can be rewritten in baseband terms as follows

\begin{equation}
\label{eq.bprs}
\mathbf{y}_{k} = \mathbf{W}_{BB_k}^H  \mathbf{H}_{BB_k} \mathbf{F}_{BB} \mathbf{s} + \mathbf{W}_{BB_k}^H  \mathbf{n}_{BB_k}
\end{equation}

\noindent where $\mathbf{H}_{BB_k} = \mathbf{W}_{RF_k}^H   \mathbf{H}_k \mathbf{F}_{RF}$ represents the equivalent baseband channel of the user $k$ and in a similar way $\mathbf{n}_{BB_k} \sim \mathcal{CN} \left( \mathbf{0}, \mathbf{K}_{BB_k}  \right)$ is the equivalent baseband noise vector, with covariance matrix $\mathbf{K}_{BB_k} =  \sigma_n^2 \mathbf{W}_{RF_k}^H  \mathbf{W}_{RF_k}$. Since the combiner and precoder matrix, $\mathbf{W}_{BB_k}$ and $\mathbf{F}_{BB}$, do not have constraints for its generation, they can be designed from signal processing approaches that take into account the colored Gaussian noise. Therefore, the problem is in the selection of the analog matrix $\mathbf{W}_{RF_k}$ and $\mathbf{F}_{RF}$, such that an equivalent baseband channel that facilitates the digital processing is obtained.

The signal-to-noise ratio (SNR) is defined as

\begin{eqnarray}
\label{eq.snr}
\mathrm{SNR} & = & \frac{\mathbb{E} [ \| \mathbf{F}_{RF}\mathbf{F}_{BB} \mathbf{s} \|^2]}{\sigma_n^2} \nonumber \\ 
 & = & \frac{Tr\left( \mathbf{F}_{RF}\mathbf{F}_{BB} \mathbb{E} [\mathbf{s} \mathbf{s}^H ] \mathbf{F}_{BB}^H \mathbf{F}_{RF}^H ] \right)}{\sigma_n^2} \nonumber \\ 
 & = & \frac{ \parallel \mathbf{F}_{RF}\mathbf{F}_{BB} \parallel_F^2}{ \sigma_n^2} =  \frac{K N_s}{ \sigma_n^2} = \frac{E_T}{\sigma_n^2}
\end{eqnarray}


\noindent where $E_T = K N_s$ represents the total energy available at the BS for transmission.

\section{Channel model}
\label{Sec.systemchannel}
The mmWave channel can be described as follows \cite{13}

\begin{eqnarray}
\mathbf{H}_k & = & \sqrt{ \dfrac{N_t N_r}{N_p}} \sum_{p=1}^{N_p} \alpha_{k,p} \mathbf{d}_{N_r}(  \mathit{h} (\phi_{k,p}^r), \mathit{v} (\theta_{k,p}^r) ) \nonumber \\
& & \qquad \qquad \qquad \qquad  \mathbf{d}_{N_t}(  \mathit{h} (\phi_{k,p}^t), \mathit{v} (\theta_{k,p}^t) )^H 
\end{eqnarray}

\noindent where $N_p$ is the number of multi-path components in the channel;  $\alpha_{k,p} \backsim \mathcal{CN}(0,1)$ is the  complex gain of the $p$-th multi-path component in the channel for the $k$-th user, whereas $\phi_{k,p}^r$ ($\theta_{k,p}^r$) and $\phi_{k,p}^t$ ($\theta_{k,p}^t$) are its azimuth (elevation) angles of arrival and departure, respectively \cite{1}.  We consider the use of an uniform planar array (UPA) formed by $N_t = N_{t_h} N_{t_v}$ ($N_r = N_{r_h} N_{r_v}$) antennas, $N_{t_h}$ ($N_{r_h}$) antennas in the horizontal side and $N_{t_v}$ ($N_{r_v}$) antennas in the vertical side, with the antenna spacing of half wave length at the transmitter (receiver)\cite{8}, whose response is given by:

\begin{equation}
\mathbf{d}_{N_t}( \mathit{h} (\phi), \mathit{v} (\theta) ) = \mathbf{d}_{N_{t_h}}( \mathit{h} (\phi)) \otimes \mathbf{d}_{N_{t_v}}( \mathit{v} (\theta))
\end{equation} 

\noindent with $\mathit{h} (\phi) = \pi cos(\phi) sin(\theta)$;  $\mathit{v} (\theta) = \pi cos(\theta)$; and

\begin{equation}
\mathbf{d}_{M}(\psi ) = \dfrac{1}{\sqrt{M}} \left[1, e^{j\psi}, ..., e^{j (M-1) \psi} \right]^T \in \mathbb{C}^{M \times 1}
\end{equation}

\section{Hybrid designing approaches}
This section presents three different approaches for designing the HB for both the transmitter and the receivers. The common factor in theses designs is that they use a HB generation divided in two stages, where the first (second) stage obtains the analog (digital) part of the precoder and of the combiner. In order to stress the main characteristic of each stage in a HB design, we considered the following notation to refer to them hereafter: [$ \cdot $]-S$_1$-S$_2$, where the first term indicates the reference number and the description of S$_1$ (S$_2$) is related to the first (second) stage-characteristic. An asterisk appearing as an upper index, $^*$, in a given characteristic means that it has been modified. In addition, the replacement of the reference number by the letter P is used to refer to our proposals.

%

\label{sec.HBapproaches}
\subsection{\cite{21}-SVD-MMSE algorithm}
The design of the HB in \cite{21} is based on channel knowledge by each user,  the analog combiner of each user is independently designed based on the singular value decomposition (SVD), while the analog precoder is obtained by  conjugate transposition to maximize the effective channel gain. Then, with the resulting equivalent baseband channel, low dimensional baseband precoders can be efficiently applied, e.g., MMSE or ZF filter. The Algorithm \ref{a.21} resumes the steps for the HB designing in \cite{21}. Note that the considered MMSE filter in the step 4 is a pseudo MMSE linear precoder. A more appropriate expression for the MMSE filter is demonstrated in the Appendix and used in the proposals described in sections \ref{sec.pro1}, \ref{sec.pro2.3} and \ref{sec.pro4}. 

\begin{algorithm}[h!]
\caption{\cite{21}-SVD-MMSE algorithm}
\label{a.21}
\begin{algorithmic}[1]
\STATE Description of the inputs and outputs\\
\textbf{Inputs}: $\mathbf{H}_k$, $k=1,...,K$\\
\textbf{Output}: $\mathbf{F}_{RF}$, $\mathbf{F}_{BB}$, $\mathbf{W}_{RF_k}$, $\mathbf{W}_{BB_k} = \mathbf{I}_{N_s}$ 
\STATE Compute the analog beamforming precoder and combiner of each user  \\
$\mathbf{W}_{RF_k} = \frac{1}{\sqrt{N_r}} \Psi \left( \mathbf{U}_{k_{:,1:N_s}} \right)$, where $\mathbf{H}_k  = \mathbf{U}_k \mathbf{\Sigma}_k \mathbf{V}_k^H $ \\
$\mathbf{F}_{RF_k} = \frac{1}{\sqrt{N_t}}  \Psi \left(  \mathbf{H}_k^H \mathbf{W}_{RF_k} \right) $\\
$\mathbf{F}_{RF} = \begin{bmatrix}
\mathbf{F}_{RF_1} & \mathbf{F}_{RF_2} & \cdots & \mathbf{F}_{RF_K}
\end{bmatrix}$
\STATE Compute the equivalent baseband channel \\
$\mathbf{H}_{BB_k} =  \mathbf{W}_{RF_k}^H \mathbf{H}_k \mathbf{F}_{RF}$\\
${\mathbf{H}_{BB}} = \begin{bmatrix}
{\mathbf{H}}_{BB_1}^T & {\mathbf{H}}_{BB_2}^T & \cdots & {\mathbf{H}}_{BB_K}^T 
\end{bmatrix}^T$
\STATE Compute the digital beamforming precoder \\
$\mathbf{F}_{BB} = {\mathbf{H}_{BB}}^H( {\mathbf{H}_{BB}} {\mathbf{H}_{BB}}^H + \sigma^2_n \mathbf{I}_{KN_s})^{-1}$
\STATE Normalize in such a way that $\parallel \mathbf{F}_{RF} \mathbf{F}_{BB} \parallel _F ^2 = KN_s$
\end{algorithmic}
\end{algorithm}

\subsection{\cite{29}-CIA-BD algorithm}
In \cite{29}, the authors focused on the design of the equivalent baseband channel, i.e., the analog part of the combiner and precoder, and eliminated the interference through baseband block diagonalization (BD) precoding. 

The analog part of the combiner is obtained through the optimization problem:

\begin{eqnarray}
\label{eq.c.cia}
\max_{ \scriptsize{ \mathbf{W}_{RF_k}} 
}{ \mathrm{det} \left( \mathbf{W}_{RF_k}^H \mathbf{A}_k \mathbf{W}_{RF_k} \right) } \\
\text{s.t. } \parallel \left(\mathbf{W}_{RF_k} \right)_{i,j} \parallel = 1/\sqrt{N_r} \: \forall i,j  \nonumber
\end{eqnarray}

\noindent where $\mathbf{A}_k = \mathbf{H}_k \mathbf{H}_k^H$. The solution of (\ref{eq.c.cia}) is reached using an column iterative  algorithm (CIA) defined in \cite{29_8} and presented in Algorithm \ref{a.column} \cite{29}.

\begin{algorithm}[h!]
\caption{Column iterative algorithm - \cite{29_8}}
\label{a.column}
\begin{algorithmic}[1]
\STATE $\mathbf{B} = column\text{-}iterative( \mathbf{D} )$
\STATE Description: this function finds a solution for \\

\hspace{30pt} $\argmax_{ \scriptsize{ \mathbf{B}} 
}{ \mid \mathbf{B}^H \mathbf{D}  \mathbf{B} \mid } $\\
\hspace{30pt} s.t. $\mid  \mathbf{B}_{i,j} \mid = 1/\sqrt{N_n}$ $\forall i,j$

where $\mathbf{B} \in \mathbb{C}^{N_n \times N_m} $ and $\mathbf{D} \in \mathbb{C}^{N_n \times N_n} $

\STATE Definitions \\
$\mathbf{\bar{B}}_{j}$ is the submatrix of $\mathbf{B} \in \mathbb{C}^{N_n \times N_m} $ with the $j$-th column vector removed\\
$\mathbf{C}_j = \mathbf{\bar{B}}_{j}^H \mathbf{D} \mathbf{\bar{B}}_{j}$\\
$\mathbf{G}_j = \mathbf{D} - \mathbf{D} \mathbf{\bar{B}}_{j}\mathbf{C}_j^{-1} \mathbf{\bar{B}}_{j}^H \mathbf{D} $\\
\STATE Optimizing \\
$\mathbf{B}^{(0)} =  \frac{1}{\sqrt{N_n}}\mathbf{1}_{N_n \times N_m}$\\
\textbf{while} $\mathbf{B}$ does not converge \\
\hspace{10pt} Update the iteration counter $k = k + 1$ \\
\hspace{10pt} \textbf{for} $j = 1,..., N_m$ \\
\hspace{20pt} Compute $ \mathbf{C}_j$ and $ \mathbf{G}_j$\\
\hspace{20pt} \textbf{for} $i = 1,..., N_n$ \\	
\hspace{30pt} $\mathbf{B}_{i,j}^{(k)} = \frac{1}{\sqrt{N_n}}\Psi \left( \sum_{l \neq i} (\mathbf{G_j})_{i,j} \mathbf{B}_{l,j}^{(k-1)}   \right)$\\
\hspace{20pt} \textbf{end} \\
\hspace{10pt}	\textbf{end} \\
\textbf{end}\\
\textbf{return} $\mathbf{B} = \mathbf{B}^{(k)}$
\end{algorithmic}
\end{algorithm}

The analog part of the precoder is obtained using the Algorithm \ref{a.column} over the objective function in (\ref{eq.p1}).

\begin{eqnarray}
\label{eq.p1}
\max_{ \scriptsize{ \mathbf{F}_{RF}} 
}{ \mathrm{det}\left( \mathbf{F}_{RF}^H \mathbf{A} \mathbf{F}_{RF} \right) } \\
\text{s.t. } \parallel  \left(\mathbf{F}_{RF} \right)_{i,j} \parallel = 1/\sqrt{N_t} \: \forall i,j  \nonumber
\end{eqnarray}

\noindent where $\mathbf{A} = \mathbf{H}^H \mathbf{W}_{RF} \mathbf{W}_{RF}^H \mathbf{H}$, with $\mathbf{W}_{RF} = \mathrm{ blkdiag} \{ \mathbf{W}_{RF_1}, ..., \mathbf{W}_{RF_K} \}$ and $\mathbf{H} = \begin{bmatrix}
\mathbf{H}_1^T & ... & \mathbf{H}_K^T
\end{bmatrix}^T$. Algorithm \ref{a.29} presents a global summary of the analog beamforming design in \cite{29}.

\begin{algorithm}[h!]
\caption{Analog beamforming design -\cite{29} algorithm}
\label{a.29}
\begin{algorithmic}[1]
\STATE Description of the inputs and outputs\\
\textbf{Inputs}: $\mathbf{H}_k$, $k=1,...,K$\\
\textbf{Output}: $\mathbf{F}_{RF}$,  $\mathbf{W}_{RF_k}$ 
\STATE Compute the analog beamforming combiner of each user \\
$\mathbf{A_k} = \mathbf{H}_k \mathbf{H}_k^H$\\
$\mathbf{W}_{RF_k} = column\text{-}iterative( \mathbf{A_k} )$\\
\STATE Computing the analog beamforming precoder  \\
$\mathbf{A} = \mathbf{H}^H \mathbf{W}_{RF} \mathbf{W}_{RF}^H \mathbf{H}$\\
$\mathbf{F}_{RF} = column\text{-}iterative( \mathbf{A} )$\\
\end{algorithmic}
\end{algorithm}

For the digital beamforming part, the BD is considered. A review of the BD precoder is presented in  Algorithm \ref{a3.BD} \cite{BDdelamare}, where the inputs are the equivalent baseband channels of the users, ${\mathbf{H}}_{BB_k}$, and outputs are $\mathbf{W}_{BB_k}$ and $\mathbf{F}_{BB}$.  In addition, a normalization constant has to be computed to satisfy the constraint $\parallel \mathbf{F}_{RF} \mathbf{F}_{BB} \parallel _F ^2 = KN_s$.
  
\begin{algorithm}[h!]
\caption{Review of the BD precoder algorithm}
\label{a3.BD}
\begin{algorithmic}[1]
\STATE Description of the inputs and outputs\\
\textbf{Inputs}: $\mathbf{H}_{BB_k}$, $k=1,...,K$\\
\textbf{Output}: $\mathbf{F}_{BB}$ and $\mathbf{W}_{BB_k}$
\STATE Definitions \\
$\mathbf{F}_{BB_k} = \mathbf{F}_k^a \mathbf{F}_k^b$ and $\mathbf{F}_{BB} = \begin{bmatrix}
\mathbf{F}_{BB_1} & \cdots & \mathbf{F}_{BB_K}
\end{bmatrix}$.

\STATE Compute the null space to avoid the multiuser interference \\
$\mathbf{\bar{H}}_{k}  =  \left[
\mathbf{H}_{BB_1}^T  \cdots  \mathbf{H}_{BB_{k-1}}^T  \mathbf{H}_{BB_{k+1}}^T  \cdots  \mathbf{H}_{BB_K}^T \right]^T $ \\
$\mathbf{\bar{H}}_k = \mathbf{\bar{U}}_k \mathbf{\bar{\Sigma }}_k\mathbf{\bar{V}}_i^H = \mathbf{\bar{U}}_k \mathbf{\bar{\Sigma }}_k \left[\mathbf{\bar{V}}_k^{(1)} \mathbf{\bar{V}}_k^{(0)} \right]^H $\\
$\mathbf{F}_k^a = \mathbf{\bar{V}}_k^{(0)} \in \mathbb{C}^{N_{RF_t} \times N_s}$ \\
Note that $\forall k \in (1,...,K)$ $\mathbf{\bar{H}}_k \mathbf{F}_k^a = \mathbf{0}$
\STATE Compute the precoder's second part to improve the energy signal as follows\\
$\mathbf{H}_{BB_k} \mathbf{F}_k^a = \mathbf{U}_k  \mathbf{\Sigma}_k\mathbf{V}_k^H = \mathbf{U}_k \mathbf{\Sigma}_k \left[\mathbf{V}_k^{(1)} \mathbf{V}_k^{(0)} \right]^H $\\
$\mathbf{F}_k^b = \mathbf{\mathbf{V}_k^{(1)}} \mathbf{\Lambda}_k$\\
where $\mathbf{\mathbf{V}_k^{(1)}}\in \mathbb{C}^{N_s \times N_s}$ and $\mathbf{\Lambda}_k$ is the user $k$'s power loading matrix that depends on the optimization criterion, e.g., waterfilling.
\STATE The user $k$'s decoding matrix is obtained as\\
$\mathbf{W}_{BB_k} = \mathbf{U}_k$
\end{algorithmic}
\end{algorithm}

\subsection{\cite{m3}-CIA-MMSE algorithm}
Our previous work in \cite{m3}, \cite{m3}-CIA-MMSE, consists in changing the digital beamforming part of \cite{29}-CIA-BD for the pseudo MMSE filter precoder defined in step 4 of the Algorithm \ref{a.21} instead of a BD filter. In this approach the hybrid combiner complexity decreases because $\mathbf{W}_{BB_k}= \mathbf{I}_{N_s}$, which means that just analog beamforming is used in the receivers.

\section{Hybrid precoder/combiner proposal I}
\label{sec.pro1}
Our first proposal, P-CIA$^{*}$-MMSE$^{*}$, is described as follows. According to \cite{29} the analog part of the combiner can be designed from the maximization of the determinant of ${\mathbf{H}}_{BB_k}{\mathbf{H}}_{BB_k}^H $, where $\mathbf{H}_{BB_k} =  \mathbf{W}_{RF_k}^H \mathbf{H}_k \mathbf{F}_{RF} $, which means consider the following problem 

\begin{eqnarray}
\label{eq.maxcapF}
\max_{ \scriptsize{ \mathbf{F}_{RF}, \mathbf{W}_{RF_k}} 
}{ \mathrm{det} \left( \mathbf{W}_{RF_k}^H \mathbf{H}_k \mathbf{F}_{RF} \mathbf{F}_{RF}^H \mathbf{H}_k^H \mathbf{W}_{RF_k} \right) } \\
\text{s.t. } \parallel  \left(\mathbf{F}_{RF} \right)_{i,j} \parallel = 1/\sqrt{N_t}, \: \parallel  \left(\mathbf{W}_{RF_k} \right)_{l,m} \parallel = 1/\sqrt{N_t} \nonumber
\end{eqnarray}

Considering an ideal case with no multiuser interference, (\ref{eq.maxcapF}) can be simplified to

\begin{eqnarray}
\max_{ \scriptsize{ \mathbf{F}_{RF_k}, \mathbf{W}_{RF_k}} 
}{ \mathrm{det} \left(  \mathbf{W}_{RF_k}^H \mathbf{H}_k \mathbf{F}_{RF_k} \mathbf{F}_{RF_k}^H \mathbf{H}_k^H \mathbf{W}_{RF_k} \right) } \\
\text{s.t. } \parallel  \left(\mathbf{F}_{RF_k} \right)_{i,j} \parallel = 1/\sqrt{N_t}, \: \parallel  \left(\mathbf{W}_{RF_k} \right)_{l,m} \parallel = 1/\sqrt{N_r} \nonumber
\end{eqnarray}

\noindent where $\mathbf{F}_{RF_k}$ is the submatrix of $\mathbf{F}_{RF}$ corresponding to the analog precoder of the user $k$. To solve this non-convex optimization problem, we can use the column iterative algorithm used in \cite{29} (see Algorithm \ref{a.column}) and to alleviate the dependence between $\mathbf{W}_{RF_k}$ and $\mathbf{F}_{RF_k}$, a recursive algorithm is considered as illustrated in Algorithm \ref{a.p.fk}.

\begin{algorithm}[h!]
\caption{Analog precoder performed by the BS}
\label{a.p.fk}
\begin{algorithmic}[1]
\STATE Description of the inputs and outputs\\
\textbf{Inputs}: $\mathbf{H}_k$, $k=1,...,K$\\
\textbf{Output}: $\mathbf{F}_{RF}$
\STATE Initial settings\\
$\mathbf{W}_{RF_k} = \frac{1}{\sqrt{N_r}}\mathbf{1}_{N_r \times N_s}$, $\mathbf{F}_{RF} = \frac{1}{\sqrt{N_t}}\mathbf{1}_{N_t \times KN_s}$ \\
\STATE Computing the analog combiner and precoder \\
\textbf{while} $\mathbf{W}_{RF_k}$ and $\mathbf{F}_{RF}$ do not converge \\
\hspace{10pt} Updating $\tilde{\mathbf{A}}_k = \mathbf{H}_k \mathbf{F}_{RF_k} \mathbf{F}_{RF_k}^H \mathbf{H}_k^H$ \\
\hspace{10pt} $\mathbf{W}_{RF_k} = column\text{-}iterative( \tilde{\mathbf{A}}_k )$\\
\hspace{10pt} Updating $\mathbf{A} = \mathbf{H}^H \mathbf{W}_{RF} \mathbf{W}_{RF}^H \mathbf{H}$\\
\hspace{10pt}$\mathbf{F}_{RF} = column\text{-}iterative( \mathbf{A} )$\\
end
\end{algorithmic}
\end{algorithm}

The procedure described in Algorithm \ref{a.p.fk} is performed by the BS. For the generation of the analog part of the combiner by the MS side, it is necessary that the receivers obtain an estimate of the product $\mathbf{H}_k \mathbf{F}_{RF_k}$ which can be simpler than obtaining an estimate of just $\mathbf{H}_k $. An option to obtain this estimate is by sending pilot symbols to a single user per time without the digital precoder part, such that the  signal vector received by the user $k$ be

\begin{equation}
\mathbf{r}_k = \mathbf{H}_k \mathbf{F}_{RF_k} \mathbf{s}_k +  \mathbf{n}_k
\end{equation}

Then the MS can compute $\tilde{\mathbf{A}}_k = \mathbf{H}_k \mathbf{F}_{RF_k} \mathbf{F}_{RF_k}^H \mathbf{H}_k^H$ and finally $\mathbf{W}_{RF_k} = column\text{-}iterative( \tilde{\mathbf{A}}_k )$. Note that considering an error-free estimation, the analog part of the precoder computed in the BS and in the MB are the same. In order not to increase the complexity of the hybrid combiner generation, the digital part of the hybrid combiner is considered as an identity matrix, i.e., $\mathbf{W}_{BB_k} = \mathbf{I}_{N_s}$.

For the digital part of the hybrid precoder the received signal vector processed by the user $k$ (see equation (\ref{eq.prs})) can be rewritten as 

\begin{equation}
\mathbf{y}_k = \dot{\mathbf{H}}_k \mathbf{F}_{RF}\mathbf{F}_{BB} \mathbf{s} + \dot{\mathbf{n}}
\end{equation}

\noindent where $\dot{\mathbf{H}}_k = \mathbf{W}_{BB_k}^H  \mathbf{W}_{RF_k}^H   \mathbf{H}_k $ and $\dot{\mathbf{n}} \sim \mathcal{CN}(\mathbf{0}, \mathbf{K}_{k}  )$ with $\mathbf{K}_{k} = \sigma_n^2\mathbf{W}_{BB_k}^H  \mathbf{W}_{RF_k}^H \mathbf{W}_{RF_k}  \mathbf{W}_{BB_k}$. For this model we propose the MMSE linear filter as derived in Appendix \ref{ap.mmse}.

\begin{equation}
\label{eq.mmse}
\mathbf{F}_{BB} =  \left(  \tilde{\mathbf{H}}^H \tilde{\mathbf{H}}  + \frac{ \gamma }{KN_s} \mathbf{F}_{RF}^H \mathbf{F}_{RF} \right)^{-1} \tilde{\mathbf{H}}^H 
\end{equation}

\noindent where $\tilde{\mathbf{H}} = \begin{bmatrix}
\tilde{\mathbf{H}}_1^T & \cdots & \tilde{\mathbf{H}}_K^T
\end{bmatrix}^T$ with $\tilde{\mathbf{H}}_k = \dot{\mathbf{H}}_k \mathbf{F}_{RF}$, and $\gamma = \mathrm{tr} \{ \mathbf{K} \}$ with $\mathbf{K} = \mathrm{blkdiag} \{ \mathbf{K}_1, ..., \mathbf{K}_K  \}$ \cite{n12}. 

\section{Hybrid precoder/combiner proposal II and III}
\label{sec.pro2.3}
Our second and third proposal consist in the improvement of the digital part of the HB \cite{21}-SVD-MMSE and \cite{m3}-CIA-MMSE, which are referred hereafter as P-SVD-MMSE$^*$ and P-CIA-MMSE$^*$, respectively. This improvement is obtained through the replacement of the pseudo MMSE defined in the step 4 of Algorithm \ref{a.21} by expression (\ref{eq.mmse}).

\section{Hybrid precoder/combiner proposal IV}
\label{sec.pro4}
As previously mentioned the problem in the hybrid processing is the selection of the analog matrix $\mathbf{W}_{RF_k}$ and $\mathbf{F}_{RF}$. Since the goal is to reduce the inter user/symbol interference and also keeping SNR large, the ideal equivalent baseband channel of each user is a diagonal matrix with large entries. Consequently, obtaining an approximation for this diagonal matrix is a good approach for the design of the analog parts. The following subsections describe our fourth proposal for the hybrid combiner/precoder construction, which is referred hereafter as P-SVD$^*$-MMSE$^*$.

\subsection{Hybrid combiner proposal}
\label{sec.mim}

Based on (\ref{eq.prs}) the mutual information between the information signal sent by the BS and the user $k$ can be written as follows 

\begin{equation}
\label{eq.i}
\mathbf{I} = \log_2 \mathrm{det} \left(  \mathbf{I}_{N_s} + \left( \sigma_n^2 \mathbf{W}_k^H \mathbf{W}_k \right)^{-1}  \mathbf{W}_k^H  \mathbf{H}_k \mathbf{FF}^H \mathbf{H}_k^H \mathbf{W}_{k}  \right)
\end{equation}

\noindent where $\mathbf{W}_k =  \mathbf{W}_{RF_k}\mathbf{W}_{BB_k}$ and $\mathbf{F} =  \mathbf{F}_{RF}\mathbf{F}_{BB}$. One approach for the designing of the hybrid combiner is to find a couple of matrix, $\mathbf{W}_{BB_k}$ and $\mathbf{W}_{RF_k}$, that maximize (\ref{eq.i}) under the hardware constraints considered in mmWave scenarios. We then consider the optimization problem expressed by

\begin{eqnarray}
\label{e.mtp}
\max_{ \scriptsize{ \mathbf{W}_{RF_k}, \mathbf{W}_{BB_k}} 
}{ \mathbf{I} } \hspace{1.5cm}  \\
\text{s.t. } \parallel \left(\mathbf{W}_{RF_k} \right)_{i,j} \parallel = 1/\sqrt{N_r}  \nonumber
\end{eqnarray}

Many methods have been proposed in the literature to solve (\ref{e.mtp}), e.g., \cite{1,29,n11}, however, these methods involve complicated mathematical developments leading to complex solutions. In our approach we consider the approximation of the mutual information for large SNR.

\begin{equation}
\label{eq.i2}
\tilde{\mathbf{I}} = \log_2 \mathrm{det} \left( \left( \sigma_n^2 \mathbf{W}_k^H \mathbf{W}_k \right)^{-1}  \mathbf{W}_k^H  \mathbf{H}_k \mathbf{FF}^H \mathbf{H}_k^H \mathbf{W}_{k}  \right)
\end{equation}

Then, using the properties of the determinant and taking (\ref{eq.snr}) into account, (\ref{eq.i2}) can be rewritten as:

\begin{equation}
\label{eq.i3}
\tilde{\mathbf{I}} = \log_2  \mathrm{det} \left(   \mathrm{SNR} \mathbf{W}_k^H  \mathbf{H}_k \bar{ \mathbf{F}}\bar{ \mathbf{F}}^H \mathbf{H}_k^H \mathbf{W}_{k}  \right)  - \log_2 \mathrm{det} \left(  \mathbf{W}_k^H \mathbf{W}_k  \right)
\end{equation}

\noindent where $\bar{ \mathbf{F}} = \frac{\mathbf{F}}{ \parallel \mathbf{F} \parallel_F }$ and $\mathrm{SNR} = \frac{E_T}{\sigma_n^2 }$. Looking for a sub-optimum manageable solution for large SNR we concentrate in the channel dependent first term of (\ref{eq.i3}) and disregard the second term. Furthermore, for practical issues the combiner design should not depend on the precoder knowledge, because the users do not have access to the whole precoder matrix, even with estimation procedures they can obtain only an estimate of their precoder part, $\mathbf{F}_k$. Therefore, we consider that $\bar{ \mathbf{F}}\bar{ \mathbf{F}}^H \propto \mathbf{I}_{N_t}$. This same assumption has been considered by others authors, e.g., \cite{29}. With these simplifying assumptions the optimization problem is formulated as

\begin{eqnarray}
\label{eq.i4}
\max_{ \scriptsize{ \mathbf{W}_{RF_k}, \mathbf{W}_{BB_k}} 
}{ \mathrm{det} \left( \mathbf{W}_k^H  \mathbf{H}_k \mathbf{H}_k^H \mathbf{W}_{k}  \right) } \hspace{0cm}  \\
\text{s.t. } \parallel \left(\mathbf{W}_{RF_k} \right)_{i,j} \parallel = 1/\sqrt{N_r}  \nonumber
\end{eqnarray}

Note that a similar expression is considered in \cite{29} (see equation (\ref{eq.c.cia})). The authors in \cite{29} proposed a solution based on an iterative algorithm, whose complexity is considerable due to the matrix inversion required in each iteration (see Algorithm \ref{a.column}). From the Hadamard's inequality which states that if an arbitrary square matrix $\mathbf{A}$ a positive definite then 

\begin{equation}
\mathrm{det} \left(  \mathbf{A} \right) \leq \prod_{i} \mathbf{A}_{i,i}
\end{equation}

\noindent with equality iff $\mathbf{A}$ is a diagonal matrix \cite{elements_theory}, it is desirable that the product $\mathbf{W}_k^H  \mathbf{H}_k \mathbf{H}_k^H \mathbf{W}_{k} $ in (\ref{eq.i4}) be a diagonal matrix with large entries. From the literature on traditional MIMO systems the above can easily be reached through single value decomposition (SVD) \cite{n10}. However, in mmWave scenarios this can not be used directly due to the constraints on the number of  RF chains. To proceed with our proposed design, let us introduce the following SVD:

\begin{equation}
[ \mathbf{V}_k ,  \mathbf{\Lambda}_k, \mathbf{V}_k^H ] =  \mathbf{H}_k  \mathbf{H}_k^H
\end{equation}

We then construct the analog part of the combiner, $\mathbf{W}_{RF_k}$, from the $N_{RF_r}$ principal eigenvectors of $\mathbf{H}_k  \mathbf{H}_k^H$ as follows

\begin{equation}
\mathbf{W}_{RF_k} = \dfrac{1}{\sqrt{N_r}} \mathbf{\Psi}((\mathbf{V}_k)_{:,1:N_{RF_r}}) 
\end{equation}

Note that $\mathbf{\Psi}((\mathbf{V}_k)_{:,1:N_{RF_r}})$ is an approximation of $(\mathbf{V}_k)_{:,1:N_{RF_r}}$ when only phase shifters are used. Then, rewriting the problem in (\ref{eq.i4}) in terms of the product between the user channel and analog combiner part, $\check{\mathbf{H}}_k = \mathbf{W}_{RF_k}^H \mathbf{H}_k$, we have 

\begin{eqnarray}
\label{eq.i5}
\max_{ \scriptsize{\mathbf{W}_{BB_k}} 
}{  \mathbf{det}\left( \mathbf{W}_{BB_k}^H \check{\mathbf{H}}_k \check{\mathbf{H}}_k^H \mathbf{W}_{BB_k} \right) } \\
\end{eqnarray}

\noindent which represents a baseband problem similar to (\ref{eq.i4}) but without constraints on the matrix construction, therefore, it can be solved using 

\begin{equation}
\label{eq.w.v.psi}
\mathbf{W}_{BB_k} = (\tilde{\mathbf{V}}_k)_{:,1:N_s} 
\end{equation}

\noindent where $ \check{\mathbf{V}}_k , \check{\mathbf{\Lambda}}_k, \check{\mathbf{V}}_k^H ] =  \check{\mathbf{H}}_k \check{\mathbf{H}}_k^H $. Algorithm \ref{a.p} summarizes the steps for the realization of the proposed hybrid combiner.

\begin{algorithm}[h!]
\caption{Proposed hybrid combiner}
\label{a.p}
\begin{algorithmic}[1]
\STATE Description of the inputs and outputs\\
\textbf{Inputs}: $\mathbf{H}_k$\\
\textbf{Output}: $\mathbf{W}_{RF_k}$, $\mathbf{W}_{BB_k}$
\STATE Compute the analog beamforming combiner \\
$[ \mathbf{V}_k ,  \mathbf{\Lambda}_k, \mathbf{V}_k^H ] =  \mathbf{H}_k \mathbf{H}_k ^H$\\
$\mathbf{W}_{RF_k} = \frac{1}{\sqrt{N_r}} \mathbf{\Psi}((\mathbf{V}_k)_{:,1:N_{RF_r}})$
\STATE Compute the digital beamforming combiner \\
$[ \check{\mathbf{V}}_k , \check{\mathbf{\Lambda}}_k, \check{\mathbf{V}}_k^H ] = \check{\mathbf{H}}_k \check{\mathbf{H}}_k^H$, where $\check{\mathbf{H}}_k = \mathbf{W}_{RF_k}^H\mathbf{H}_k $.\\
$\mathbf{W}_{BB_k} = \check{\mathbf{V}}_{:,1:KN_s} $\\

\end{algorithmic}
\end{algorithm}

\subsection{Hybrid precoder proposal}

Our hybrid precoder design approach is divided in two parts, analog and digital. As stated in the previous subsection, the construction of the combiner's (precoder's) analog part  based on the analog approximation to the eigenvectors of the channel user (entire channel) can benefit to diagonalization of the equivalent baseband user channel (entire equivalent baseband channel), which reduces the intersymbol (user) interference when is applied in the MS (BS). Thus, for the analog part of the precoder we consider the mutual information maximization problem of the entire channel,  $\mathbf{H} = \begin{bmatrix} \mathbf{H}_{1}^T & \cdots & \mathbf{H}_{K}^T \end{bmatrix}^T$, when only analog processing is available in the transmitter, and we solve it through the same methodology used for the hybrid combiner realization.
  
Considering only analog processing in the transmitter with large SNR, the mutual information maximization problem of the entire channel can be reduced to the following problem

\begin{eqnarray}
\max_{ \scriptsize{ \mathbf{F}_{RF}} 
}{ \mathrm{det}\left(  \mathbf{W}^H\mathbf{H} \mathbf{F}_{RF} \mathbf{F}_{RF}^H  \mathbf{H}^H  \mathbf{W} \right) } \hspace{0.5cm}  \\
\text{s.t. }  \parallel  \left(\mathbf{F}_{RF} \right)_{i,j} \parallel = 1/\sqrt{N_t}  \nonumber
\end{eqnarray}

\noindent or equivalently as\footnote{Consider an arbitrary complex matrix $\mathbf{A} \in \mathcal{C}^{n \times m}$ with $n\leq m$. If $\mathbf{A}\mathbf{A}^H $ is full ranking, then $\mathrm{det}\left( \mathbf{A}\mathbf{A}^H \right) = \prod_{i=1}^{n} \lambda_i = \mid \mathbf{A}^H\mathbf{A} \mid$, where $\lambda_i$ is the $i$-th eigenvalue of $\mathbf{A}\mathbf{A}^H$.}.

\begin{eqnarray}
\max_{ \scriptsize{ \mathbf{F}_{RF}} 
}{ \mid \mathbf{F}_{RF}^H  \mathbf{H}^H  \mathbf{W} \mathbf{W}^H\mathbf{H} \mathbf{F}_{RF} \mid  } \hspace{0.5cm}  \\
\text{s.t. }   \parallel  \left(\mathbf{F}_{RF} \right)_{i,j}  \parallel = 1/\sqrt{N_t}  \nonumber
\end{eqnarray}

\noindent where $\mathbf{W} = \mathrm{blkdiag} \{ \mathbf{W}_1, ..., \mathbf{W}_K \}$. To construct $\mathbf{F}_{RF}$ we use the same methodology of the hybrid combiner described in Subsection \ref{sec.mim}. Therefore, the suboptimum analog part of the precoder is obtained as

\begin{equation}
\label{eq.FRF}
\mathbf{F}_{RF} = \frac{1}{\sqrt{N_t}} \mathbf{\Psi}  \left(\mathbf{V}_{:,1:N_{RF_t}} \right)
\end{equation}

\noindent where $[\mathbf{V}, \mathbf{\Lambda}, \mathbf{V}^H ] =  \mathbf{H}^H  \mathbf{W} \mathbf{W}^H \mathbf{H}  $. 

For the digital part of the hybrid precoder, low dimensional linear filters can be used, we considered the MMSE filter described in our first proposal, i.e., $\mathbf{F}_{BB}$ is given by (\ref{eq.mmse}). With $\mathbf{F}_{RF}$ given by (\ref{eq.FRF}) and $\mathbf{F}_{BB}$ obtained by (\ref{eq.mmse}), the product $\mathbf{F}_{RF} \mathbf{F}_{BB}$ is further normalized such that $\parallel \mathbf{F}_{RF} \mathbf{F}_{BB} \parallel^2_F = KN_s$


%
%
%
 

\section{Data detection approaches}
\label{sec.detection}
This section presents four sub-optimal approaches to obtain $\mathbf{\hat{s}}_k$ by each user requiring different levels of parameter knowledge (or estimation) as follows:

\begin{itemize}
\item \textit{Minimum distance detection (MDD)} 
\begin{equation}
\mathbf{\hat{s}}_k = \argmin_{ \scriptsize{ \mathbf{d} \in \mathbb{Q} ^{N_s \times 1}}}{\parallel {\mathbf{y}}_k - \mathbf{A}_k\mathbf{d}  \parallel^2 }
\end{equation}
where $\mathbf{A}_k = \mathbf{W}_{BB_k}^H \mathbf{W}_{RF_k}^H \mathbf{H}_{k} \mathbf{F}_{HB_k}$, and $\mathbf{F}_{HB_k}$ is the, unknown to the receiver, submatrix of $\mathbf{F}_{RF} \mathbf{F}_{BB}$ corresponding to the hybrid precoder of user $k$.

\item \textit{Approximate MDD (AMDD)} (assumes that $\mathbf{A}_k \approx \mathbf{I}(N_s)$)
\begin{equation}
\mathbf{\hat{s}}_k = \argmin_{ \scriptsize{ \mathbf{d} \in \mathbb{Q} ^{N_s \times 1}}}{\parallel \tilde{\mathbf{y}}_k - \mathbf{d}  \parallel^2 }
\end{equation}

\item \textit{Noise whitening operation followed by MDD (NWMDD)}
\begin{equation}
\label{eq.mlwithKn}
\mathbf{\hat{s}}_k = \argmin_{\scriptsize{ \mathbf{d} \in \mathbb{Q} ^{N_s \times 1}}}{\parallel \mathbf{K}_k^{-1/2} \left( {\mathbf{y}}_k - \mathbf{A}_k \mathbf{d}  \right) \parallel^2 }
\end{equation}

where $\mathbf{K}_k = \sigma_n^2 \mathbf{W}_{k}^H \mathbf{W}_{k}$.

\item \textit{Noise whitening operation followed by approximate MDD (NWAMDD)}
\begin{equation}
\mathbf{\hat{s}}_k = \argmin_{ \scriptsize{ \mathbf{d} \in \mathbb{Q} ^{N_s \times 1}}}{\parallel \mathbf{K}_k^{-1/2} \left( {\mathbf{y}}_k - \mathbf{d}  \right) \parallel^2 }
\end{equation}

\item \textit{Noise and interference whitening operation followed by MDD (NWIMDD)}
\begin{equation}
\label{eq.mlwithKn}
\mathbf{\hat{s}}_k = \argmin_{\scriptsize{ \mathbf{d} \in \mathbb{Q} ^{N_s \times 1}}}{\parallel \mathbf{K}_k^{-1/2} \left( {\mathbf{y}}_k - \mathbf{A}_k \mathbf{d}  \right) \parallel^2 }
\end{equation}

where $\mathbf{K}_{k} = \sigma_n^2 \mathbf{W}_k^H \mathbf{W}_k +  \sum_{ \substack{j=1 \\ j\neq k}}^K \mathbf{W}_k^H\mathbf{H}_k \mathbf{F}_j \mathbf{F}_j^H \mathbf{H}_k^H \mathbf{W}_k$. 

\end{itemize}

\section{Numerical results}
\label{Sec.numerical}

In the simulations, the users’ channels are generated with $N_p = 8$ multi-paths components, the azimuth and elevation departure angles values are given by a random variable with uniform distribution in the interval of (0;$2\pi$) and (0;$\pi$), respectively. The UPAs have square formats for both transmitter and receivers, i.e., $N_{t_h} = N_{t_v} = \sqrt N_t$ and $N_{r_h} = N_{r_v} = \sqrt N_r$. The maximum allowed setting for RF chains number is used for both the BS and for each MS, so that $N_{RF_t} = KN_s$ and $N_{RF_r} = Ns$. The results are averaged over $10^5$ channels generations for each user. 

\begin{figure}[h!]
\centering
\includegraphics[width=\textwidth]{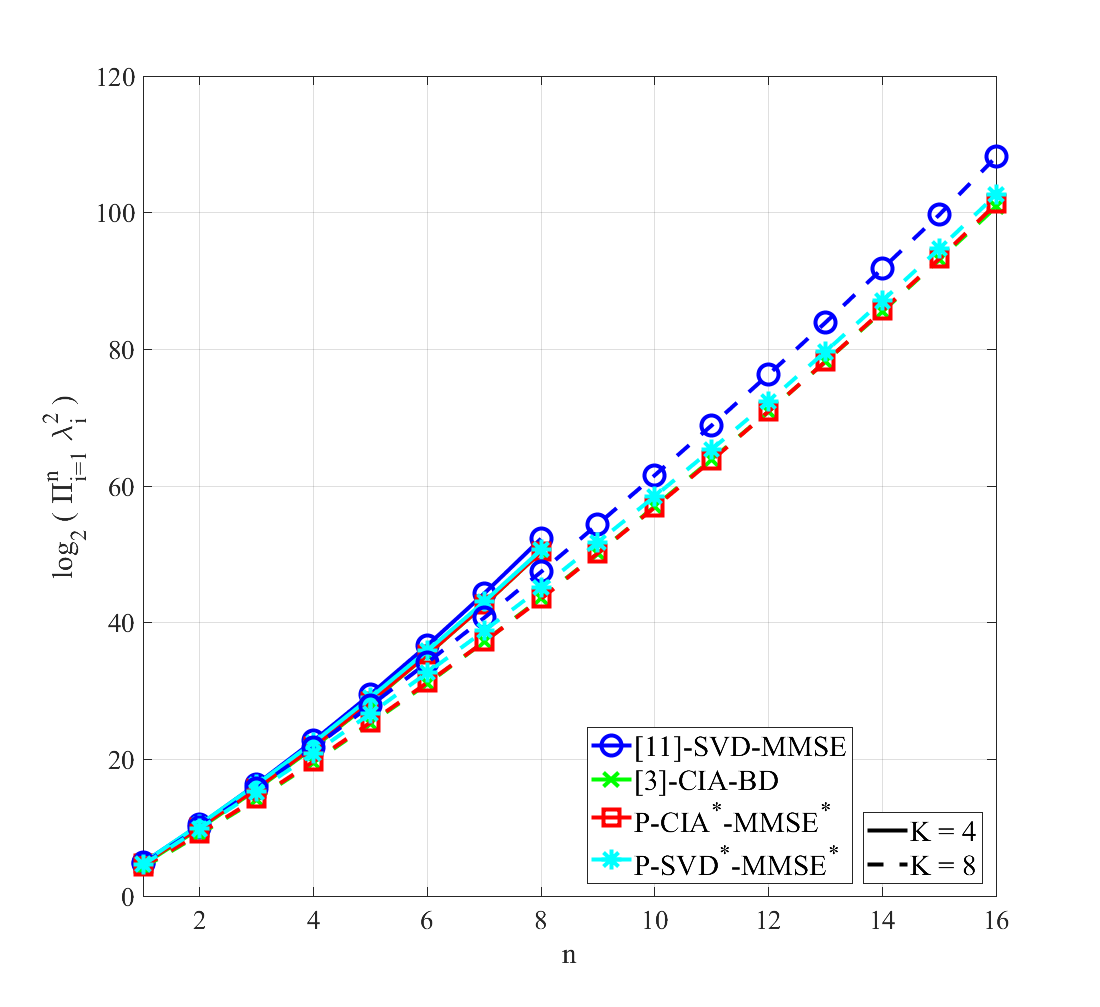}
\caption{The value of $\text{log}_2 \left( \prod_{i=1}^{n} \lambda_i^2 \right)$, with $\lambda_1 \geq \lambda_2 \geq \cdots \geq \lambda_{KN_s}  $, of different equivalent baseband channels}
\label{fig.1}
\end{figure}

Reference \cite{29} proposed an optimization involving the effective baseband user channel, $\mathbf{H}_{BB_k} = \mathbf{W}_{RF_k}\mathbf{H}_k \mathbf{F}_{RF} $, to find the analog part of the combiner/precoder. Their authors used the value of  $\log_2 \left( \prod_{i=1}^{KN_s} \lambda_i^2   \right)$, where $\lambda_i$ is the $i$-th single value of the entire effective baseband channel, $\mathbf{H}_{BB} = \begin{bmatrix}
\mathbf{H}_{BB_1}^T & \cdots & \mathbf{H}_{BB_K}^T
\end{bmatrix} ^T$, as the metric to ilustrate the advantages of their proposal. Figure \ref{fig.1} presents a comparison using this metric corresponding to the $\mathbf{H}_{BB}$ obtained by \cite{21,29} and our proposals. The simulation settings are $N_t = 64$, $N_r = 4$, $N_s = 2$ and $K=4$ ($K=8$).


From Figure \ref{fig.1} it can be observed that \cite{21}-SVD-MMSE and the hybrid designs proposed here yield the largest eigenvalues of the entire effective baseband channel. Despite the high complexity of P-CIA$^{*}$-MMSE$^{*}$ there is not a relevant gain if this metric is considered as a measure of performance.

\begin{figure}[h!]
\centering
\includegraphics[width=\textwidth]{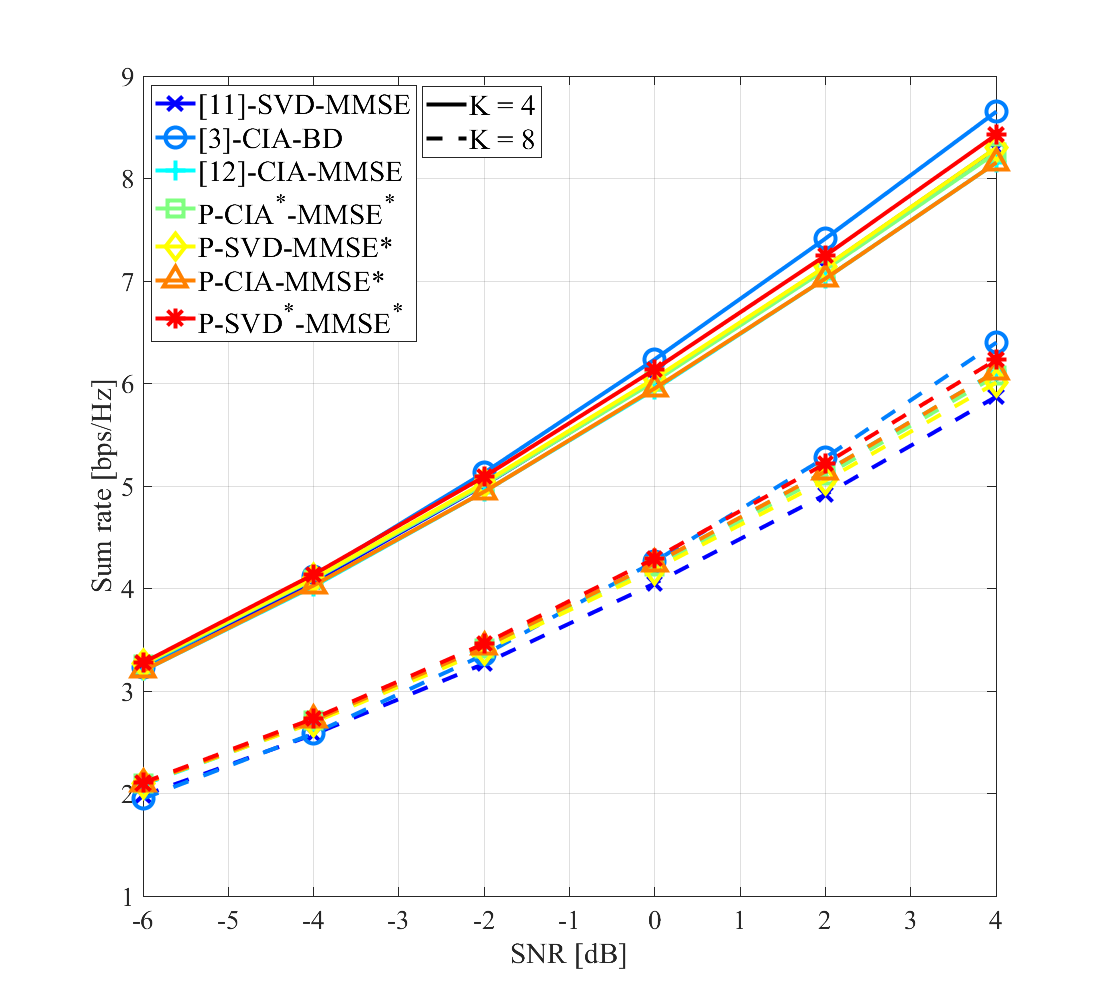}
\caption{Achievable sum rate using $N_t = 64$, $N_r = 4$, $N_s = 2$, $K = 4$ and $K=8$ }
\label{fig.2}
\end{figure}

Figure \ref{fig.2} shows a comparison in terms of sum rate computed by the expression

\begin{equation}
\sum_{k=1}^K\log_2 \mathrm{det}\left( \mathbf{I}_{N_s} + \mathbf{W}_k^H  \mathbf{H}_k \mathbf{F}_k \mathbf{F}_k^H \mathbf{H}_k^H \mathbf{W}_k \mathbf{K}_{k}^{-1} \right)
\end{equation}

\noindent where $\mathbf{K}_{k} = \sigma_n^2 \mathbf{W}_k^H \mathbf{W}_k +  \sum_{ \substack{j=1 \\ j\neq k}}^K \mathbf{W}_k^H\mathbf{H}_k \mathbf{F}_j \mathbf{F}_j^H \mathbf{H}_k^H \mathbf{W}_k$. In this figure the BS has $N_t = 64$ antennas and sends $N_s = 2$ streams to $K = 4$ and $K=8$ users equipped with $N_r = 4$.

\begin{figure}[h!]
\centering
\includegraphics[width=\textwidth]{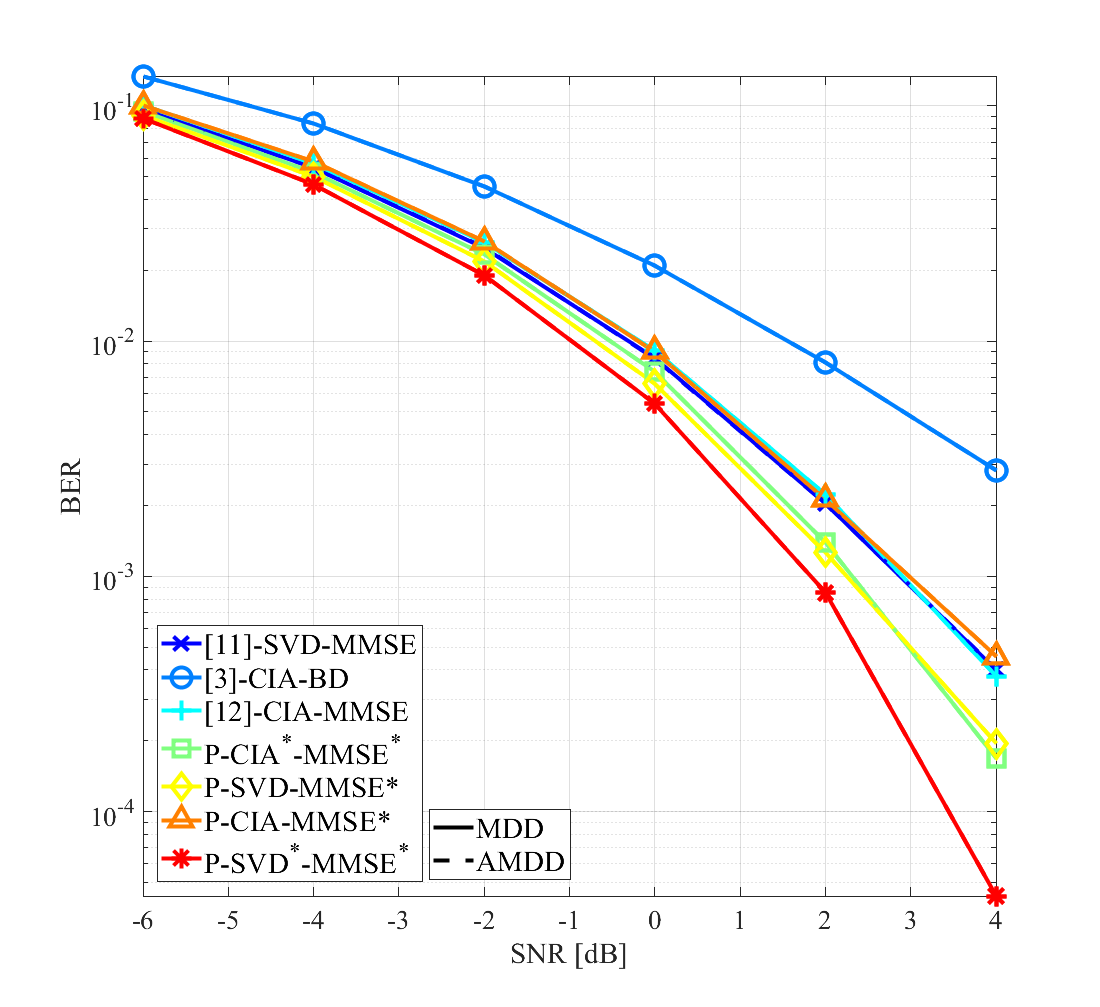}
\caption{BER results obtained with the MDD and AMDD data detectors and simulation settings $N_t = 64$, $N_r = 4$, $N_s = 2$, $K = 4$}
\label{fig.3}
\end{figure}

\begin{figure}[h!]
\centering
\includegraphics[width=\textwidth]{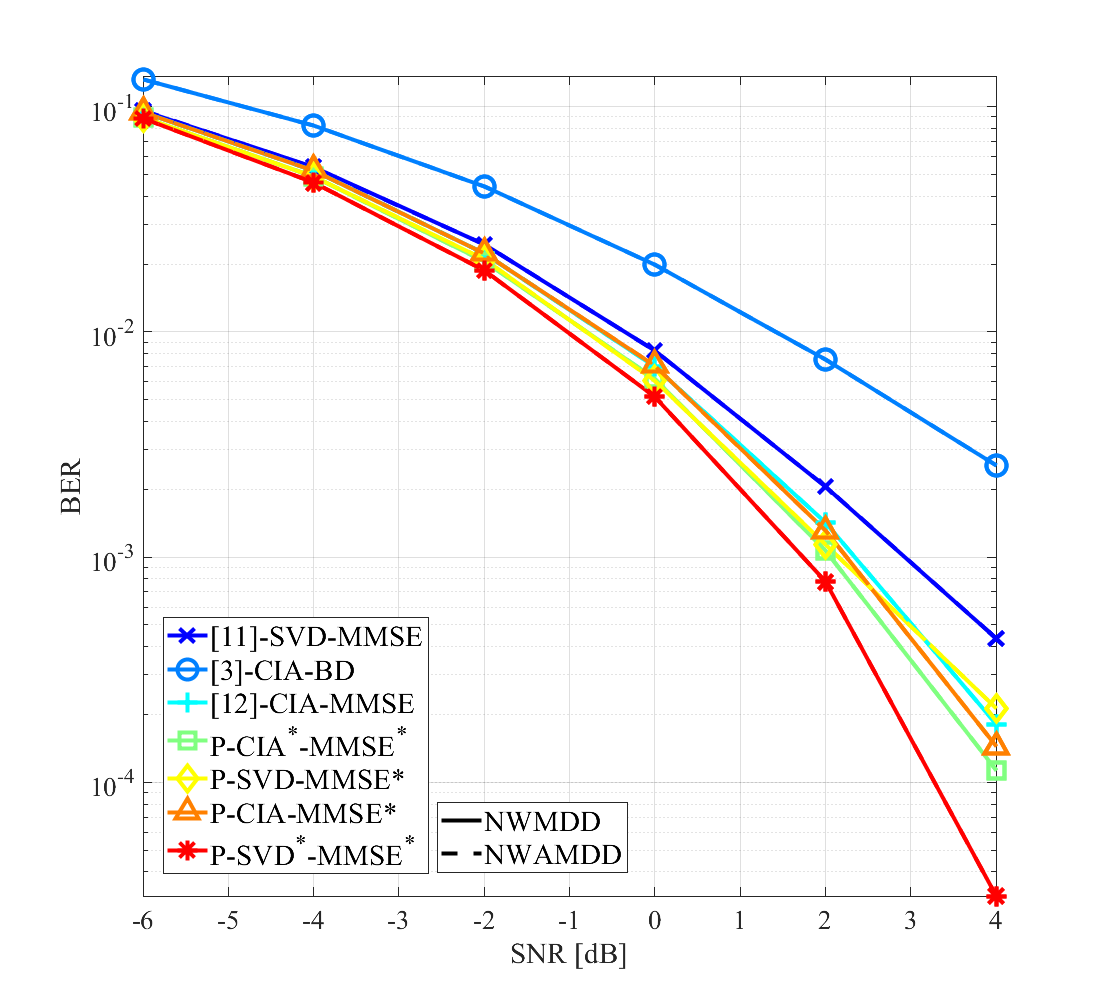}
\caption{BER results obtained with the NWMDD and NWAMDD data detectors and simulation settings $N_t = 64$, $N_r = 4$, $N_s = 2$, $K=4$ }
\label{fig.4}
\end{figure}

\begin{figure}[h!]
\centering
\includegraphics[width=\textwidth]{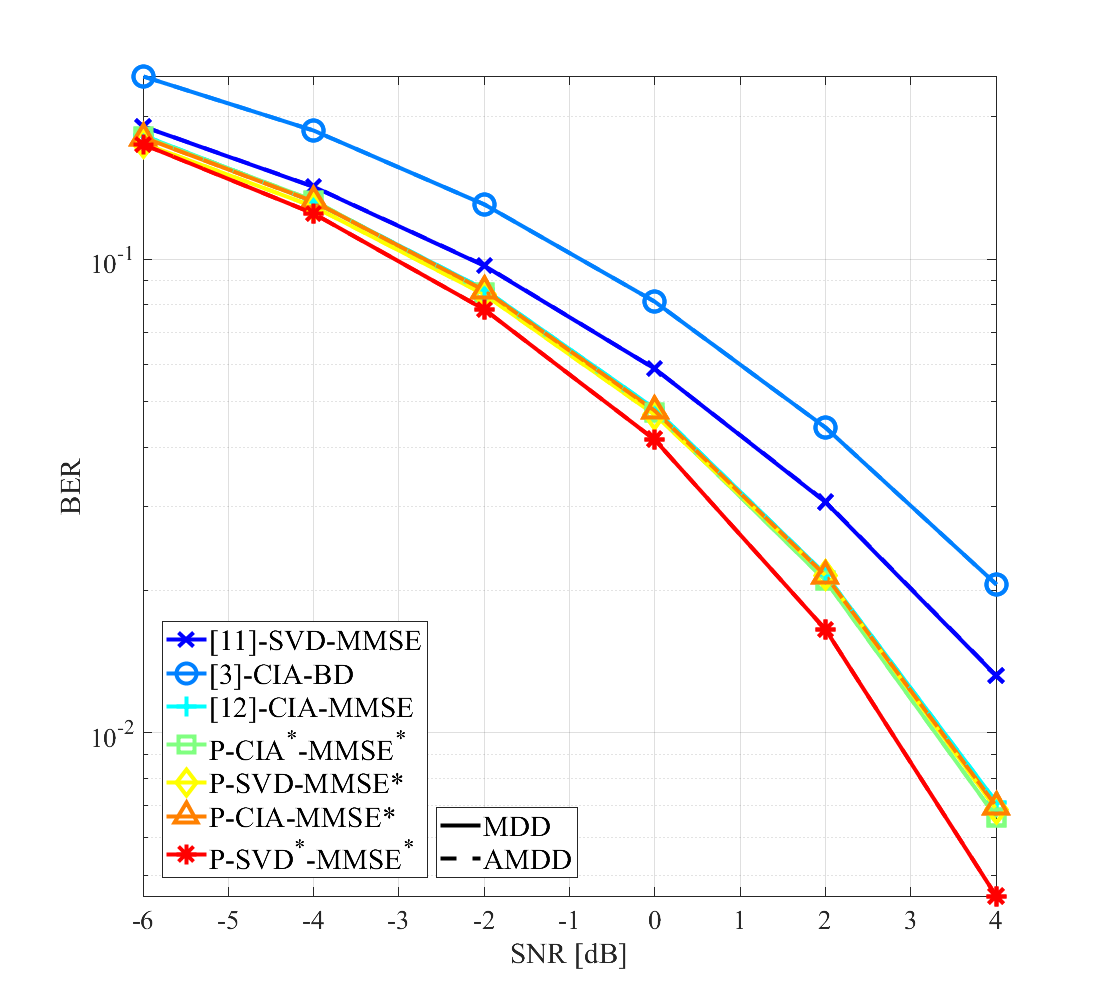}
\caption{BER results obtained with the MDD and AMDD data detectors and simulation settings $N_t = 64$, $N_r = 4$, $N_s = 2$, $K = 8$}
\label{fig.5}
\end{figure}

\begin{figure}[h!]
\centering
\includegraphics[width=\textwidth]{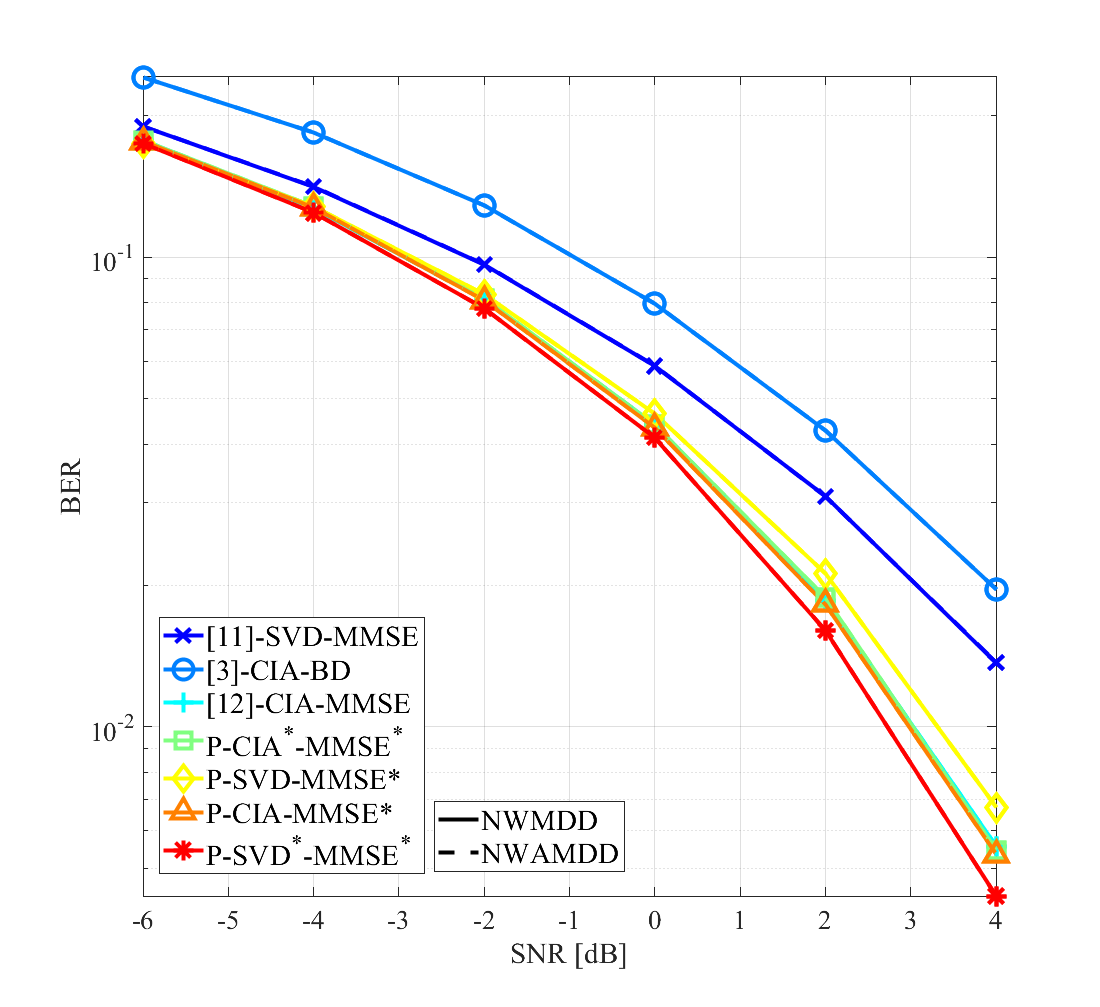}
\caption{BER results obtained with the NWMDD and NWAMDD data detectors and simulation settings $N_t = 64$, $N_r = 4$, $N_s = 2$, $K=8$ }
\label{fig.6}
\end{figure}

From Figure \ref{fig.2}, it can be observed that our proposal P-CIA$^{*}$-MMSE$^{*}$ reaches the second-best sum rate performance for the two considered scenarios, while \cite{29}-CIA-BD is the best one. In order to show the advantages of the proposed schemes in terms of BER performance we consider the following scenarios: in figures \ref{fig.3} and \ref{fig.4} the BS has $N_t = 64$ antennas and sends $N_s = 2$ streams to $K = 4$ users, each equipped with $N_r = 4$. In figures \ref{fig.5} and \ref{fig.6} the BS has $N_t = 64$ antennas and sends $N_s = 2$ streams to $K = 8$ users equipped with $N_r = 4$.

From the above mentioned figures it can be observed that the proposed P-SVD$^*$-MMSE$^*$ achieved the lowest BER values in all considered scenarios and a simple data estimation, the AMDD, can be used without a relevant performance loss. Since mmWave systems is a promising solution to short distance communications, simulations involving a number of users higher than 8 were not considered in this work.

\begin{figure}[h!]
\centering
\includegraphics[width=\textwidth]{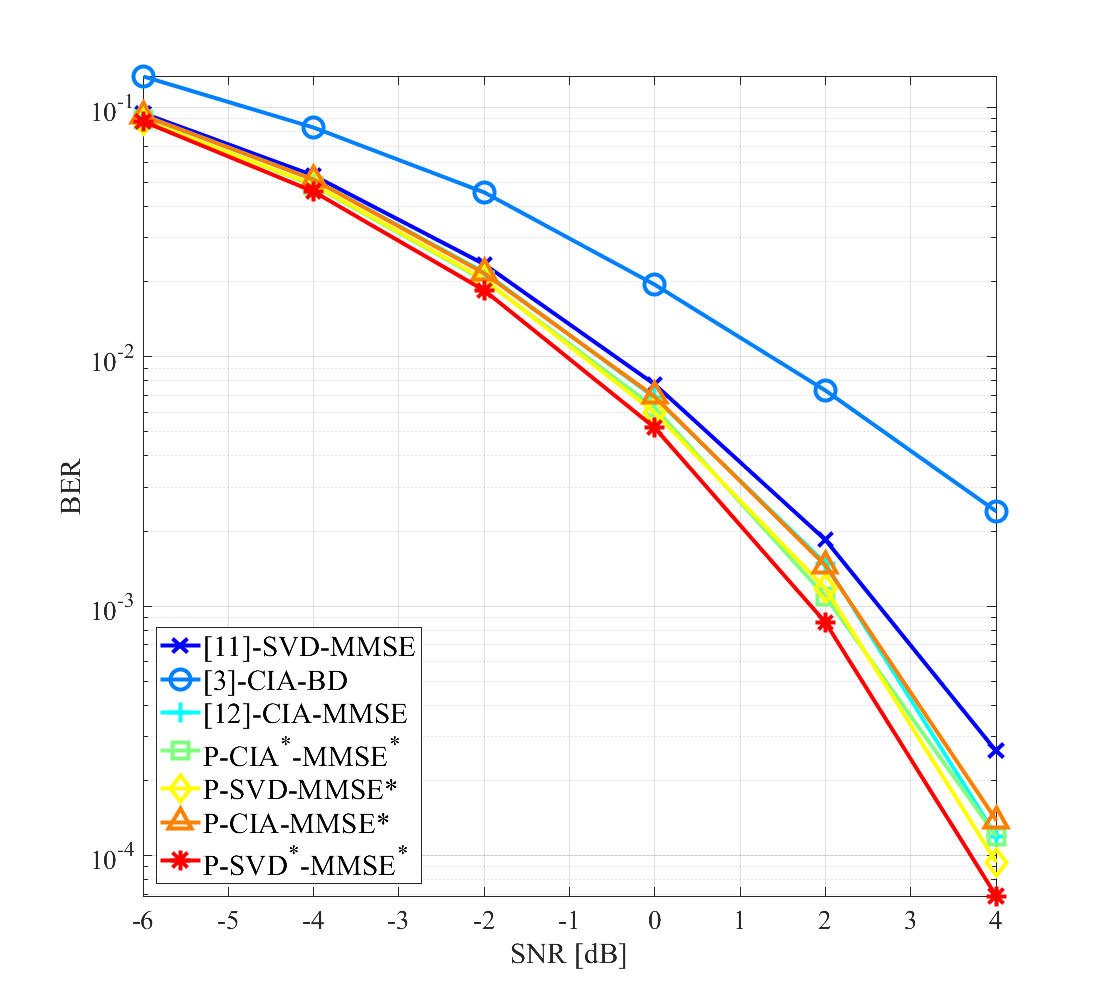}
\caption{BER results obtained with the NWIMDD data detector and simulation settings $N_t = 64$, $N_r = 4$, $N_s = 2$, $K=4$ }
\label{fig.7}
\end{figure}

\begin{figure}[h!]
\centering
\includegraphics[width=\textwidth]{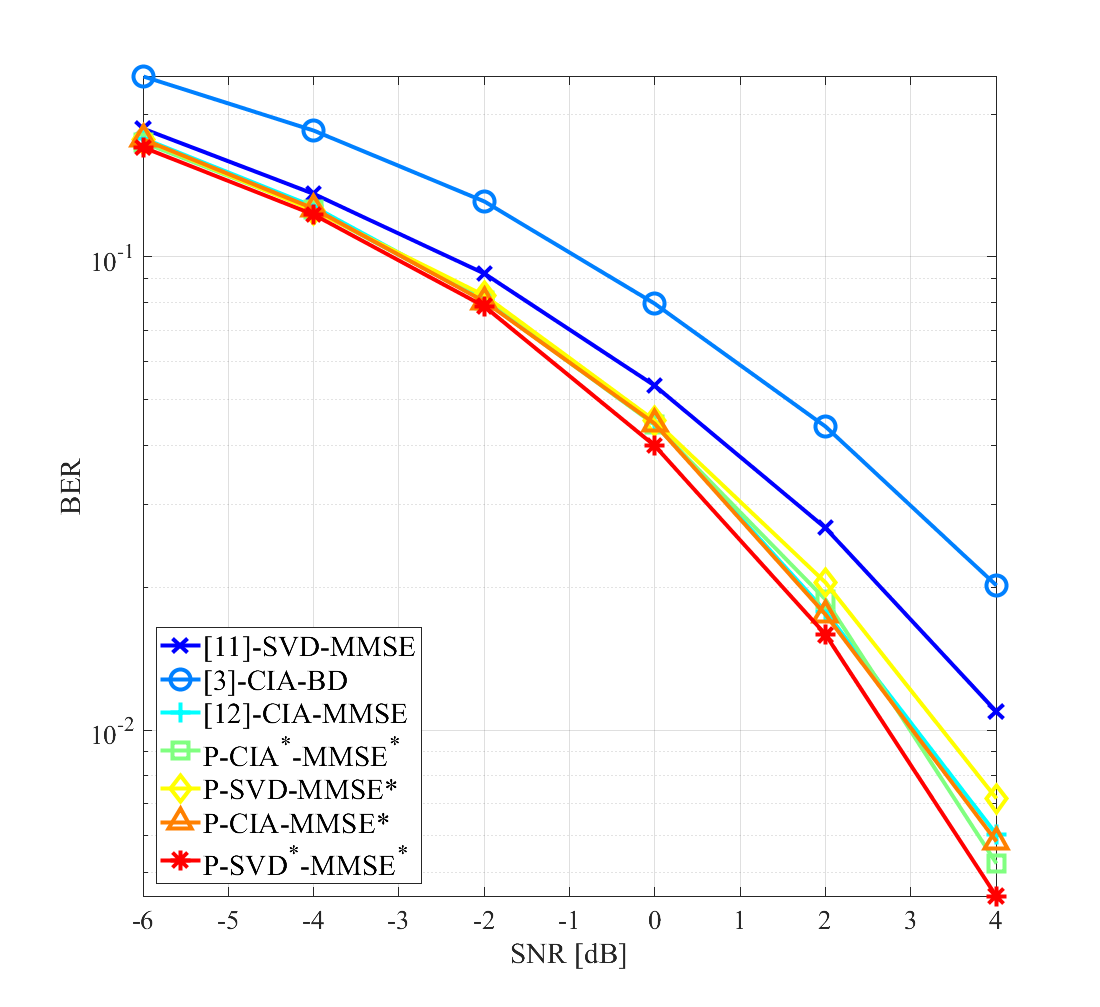}
\caption{BER results obtained with the NWIMDD data detector and simulation settings $N_t = 64$, $N_r = 4$, $N_s = 2$, $K=8$ }
\label{fig.8}
\end{figure}

Finally, Figures \ref{fig.7} and \ref{fig.8} illustrate the performance benchmark achieved by using NWIMDD in the terminals over the considered scenarios. 

\section{Conclusions}
\label{Sec.conclusions}
We proposed four hybrid combiner/precoder for downlink mmWave massive MU-MIMO systems. The designing of a hybrid combiner/precoder is divided in two parts, analog and digital. Previous designs of the analog part of the combiner/precoder are focused in the increasing of the SNR, since theses analog parts are not very useful to manage the inter user/symbol interference. A new approach to the designing of these analog parts was proposed. We noted that the ideal equivalent baseband user channel is a diagonal matrix with large entries, and the proposed designs aimed at providing an approximation to this matrix. 

Also, since the digital part of the combiner and precoder do not have constraints for their construction, theirs designs can be based on traditional signal processing that takes into account the colored noise vector in the received signal.In this respect, this paper presents a exact derivation of the MMSE linear precoder to be used as the digital part.  Numerical results illustrate the BER performance improvement obtained through the proposed hybrid designs. In addition, for the design that resulted in the best performance, P-SVD$^*$-MMSE$^*$, a simple detection approach can be used to estimate the transmitted data without a significant performance loss.

\appendices
\section{General linear constrained MMSE Precoding}
\label{ap.mmse}
Here we consider that the MIMO system consists of a linear precoder with precoding matrix $\mathbf{F}$ at the transmitter, which is constrained by a complex matrix $\mathbf{B}$. The output signal of the detector is described by
\begin{align}
\tilde{ \mathbf{x} }  =  \mathbf{{H}} \mathbf{B} \mathbf{F} \mathbf{x} + \mathbf{n} \textrm{,}
\end{align}
where $\mathbf{{H}}$ is the MIMO channel, $\mathbf{x}$ is the input signal and $\mathbf{n}$ is the additive Gaussian noise vector.
The input signal $\mathbf{x}$ has covariance matrix $\mathbb{E}\left[ \mathbf{x} \mathbf{x}^H  \right]= \mathbf{R}_{\mathbf{x}} $.
The noise $\mathbf{n}$ has covariance matrix $\mathbb{E}\left[  \mathbf{n} \mathbf{n}^H  \right]= \mathbf{R}_{\mathbf{n}} $. Because the transmit energy is constrained, the received signal is scaled by a factor $\beta$ at the receiver, which is part of the optimization.

The MMSE precoder is obtained from the following optimization problem

\begin{eqnarray} \label{eq.p}
\{\mathbf{F}_{MMSE}, \beta_{MMSE} \} = \argmin_{ \mathbf{F}, \beta } \mathbb{E}[ \parallel \mathbf{x} - \beta \tilde{\mathbf{x}} \parallel^2] \\
\text{s.t. } \mathbb{E}[ \parallel \mathbf{BFx} \parallel^2] = \mathrm{tr} \left\{  \mathbf{A}  \mathbf{F} \mathbf{R}_x \mathbf{F}^H   \right\} \leq E_{T}, \nonumber 
\end{eqnarray}

\noindent where $\mathbf{A} = \mathbf{B}^H \mathbf{B}$ and $E_T$ is the total transmit energy. The necessary conditions for the precoder filter $\mathbf{F}$ and weight $\beta \in \mathbb{R}_+$ can be found by constructing the Lagrangian function

\begin{align}
L \left(\mathbf{F}, \beta, \lambda  \right) = & \mathbb{E}[ \parallel \mathbf{x} - \beta  \tilde{\mathbf{x}} \parallel^2] + \lambda \left( \mathrm{tr}\left\{  \mathbf{A}  \mathbf{F} \mathbf{R}_x \mathbf{F} \right\} - E_T\right) \\ \notag
 = &  \mathrm{tr}\left\{ \mathbf{R}_{x} \right\}  -  \beta \ \mathbb{E}\left[    \mathbf{x}^H \mathbf{\bar{H}}\mathbf{F} \mathbf{x}        \right] 
-  \beta \ \mathbb{E}\left[    \mathbf{x}^H   \mathbf{F}^H  \mathbf{\bar{H}}^H   \mathbf{x}        \right] \\ \notag
& + \beta^2 \ \mathbb{E}\left[ \mathbf{x}^H   \mathbf{F}^H  \mathbf{\bar{H}}^H      \mathbf{\bar{H}}\mathbf{F}\mathbf{x} 
     \right]  + \beta^2 \ \mathbb{E}\left[ \mathbf{n}^H   \mathbf{n}      \right]  \\ \notag
&  + \lambda \left( \mathrm{tr}\left\{  \mathbf{A}  \mathbf{F} \mathbf{R}_x \mathbf{F}^H \right\} - E_T\right)  \textrm{.}
\end{align}

\noindent where $\mathbf{\bar{H}} = \mathbf{HB}$ and $\lambda$ is the Lagrangian multiplier. By making use of the trace operator and its properties, the Lagrangian function can be rewritten as follows

\begin{align}
L \left(\mathbf{F}, \beta, \lambda  \right) = & \mathrm{tr}\left\{ \mathbf{R}_{x} \right\}  -  \beta \ \mathrm{tr}\left\{   \mathbf{\bar{H}}\mathbf{F} \mathbf{R}_x        \right\}
-  \beta \ \mathrm{tr}\left\{  \mathbf{\bar{H}}^H  \mathbf{R}_x  \mathbf{F}^H  \right\} \\ \notag
& + \beta^2 \ \mathrm{tr}\left\{  \mathbf{\bar{H}}^H  \mathbf{\bar{H}}\mathbf{F}\mathbf{R}_x \mathbf{F}^H 
     \right\}  + \beta^2 \ \mathrm{tr}\left\{    \mathbf{R}_n    \right\}  \\ \notag
&  + \lambda \left( \mathrm{tr}\left\{  \mathbf{A}  \mathbf{F} \mathbf{R}_x \mathbf{F}^H \right\} - E_T\right) \textrm{.}
\end{align}

\noindent Then, setting its derivatives to zero 

\begin{align}  \label{eq.dF}
\frac{ \partial L\left( \mathbf{F}, \beta, \lambda  \right)}{\partial \mathbf{F} ^*} =  & 2\beta^2 \ \mathbf{\bar{H}}^H  \mathbf{\bar{{H}}}\mathbf{F} \mathbf{R}_x  - 2 \beta \ \mathbf{\bar{H}}^H  \mathbf{R}_x + 2 \lambda \mathbf{A} \mathbf{F}\mathbf{R}_x = \mathbf{0}
\end{align}

\begin{align}   \label{eq.db} 
\frac{ \partial L\left( \mathbf{F}, \beta, \lambda  \right)}{\partial \beta } = & - \mathrm{tr}\left\{  \mathbf{\bar{H}}\mathbf{F} \mathbf{R}_x        \right\}
-   \ \mathrm{tr}\left\{  \mathbf{\bar{H}}^H    \mathbf{R}_x  \mathbf{F}^H  \right\} \\ \notag & + 2\beta \ \mathrm{tr}\left\{  \mathbf{\bar{H}}^H          \mathbf{\bar{H}}\mathbf{F}\mathbf{R}_x \mathbf{F}^H 
     \right\}  + 2\beta \ \mathrm{tr}\left\{    \mathbf{R}_n   \right\}   \\ \notag
     = & - 2 \ \mathrm{tr}\left\{  \mathbf{\bar{H}}^H  \mathbf{R}_x  \mathbf{F}^H  \right\} \\ \notag & + 2\beta \ \mathrm{tr}\left\{  \mathbf{\bar{H}}^H   \mathbf{\bar{H}}\mathbf{F}\mathbf{R}_x \mathbf{F}^H 
     \right\}  + 2\beta \ \mathrm{tr}\left\{ \mathbf{R}_n \right\} = 0
\end{align}

\noindent We obtain from (\ref{eq.dF}) that

\begin{equation} \label{eq.dF1}
 \beta^{-1}\ \mathbf{\bar{H}}^H  = \left( \mathbf{\bar{H}}^H  \mathbf{\bar{H}}+ \lambda \beta ^{-2} \mathbf{A} \right) \mathbf{F}
\end{equation}

\noindent thus resulting for the precoder filter the structure

\begin{equation} \label{eq.sp}
\mathbf{F} = \beta^{-1} \ \left( \mathbf{\bar{H}}^H \mathbf{\bar{H}}+ \lambda \beta ^{-2} \mathbf{A} \right)^{-1} \mathbf{\bar{H}}^H 
\end{equation}

\noindent Multiplying from the right $2 \beta \ \mathbf{R}_x \mathbf{F}^H$ in (\ref{eq.dF1}) and using the trace operator yields

\begin{align} \label{eq.db1}
2 \ \mathrm{tr} \left\{ \mathbf{\bar{H}}^H \mathbf{R}_x \mathbf{F}^H \right\}  = &  2\beta \ \mathrm{tr} \left\{ \left( \mathbf{\bar{H}}^H \mathbf{\bar{H}}+ \lambda \beta ^{-2} \mathbf{A} \right) \mathbf{F} \mathbf{R}_x \mathbf{F}^H  \right\} \\
 = &  2\beta \ \mathrm{tr} \left\{ \mathbf{\bar{H}}^H \mathbf{\bar{H}}\mathbf{F} \mathbf{R}_x \mathbf{F}^H  \right\}  + 2 \lambda \beta ^{-1} \ \mathrm{tr} \left\{  \mathbf{A}  \mathbf{F} \mathbf{R}_x \mathbf{F}^H  \right\} \nonumber
\end{align} 

\noindent Using (\ref{eq.db}) in (\ref{eq.db1}) we have that

\begin{equation} \label{eq.c0}
\lambda \beta ^{-2} = \frac{\mathrm{tr} \left\{  \mathbf{R}_n   \right\}}{ \mathrm{tr} \left\{  \mathbf{A}  \mathbf{F} \mathbf{R}_x \mathbf{F}^H \right\}} 
\end{equation}

From (\ref{eq.c0}), it is clearly evidenced that $\lambda$ takes values grater than zero, and therefore the solution of $\mathbf{F}$ is in the border of the admissible search space, then we have that equality holds in (\ref{eq.p}), $\mathrm{tr} \left\{  \mathbf{A}  \mathbf{F} \mathbf{R}_x \mathbf{F}^H   \right\} = E_{T}$, and   

\begin{equation} \label{eq.c}
\lambda \beta ^{-2}  = \frac{\mathrm{tr} \left\{  \mathbf{R}_n   \right\}}{E_T}
\end{equation}

\noindent Substituting (\ref{eq.c}) in (\ref{eq.sp}) yields

\begin{equation} \label{eq.sp1}
\mathbf{F} = \beta^{-1} \ \left( \mathbf{\bar{H}}^H \mathbf{\bar{H}}+ \frac{\mathrm{tr} \left\{   \mathbf{R}_n   \right\}}{E_T} \mathbf{A} \right)^{-1} \mathbf{\bar{H}}^H
\end{equation}


\begin{equation}
\beta^{-2} \ \mathrm{tr} \left\{ \mathbf{A} \left(  \mathbf{\bar{H}}^H \mathbf{\bar{H}}+ \frac{\mathrm{tr} \left\{  \mathbf{R}_n   \right\}}{E_T}  \right)^{-1}  \mathbf{\bar{H}}^H  \mathbf{R}_x  \mathbf{\bar{H}} \left(  \mathbf{\bar{H}}^H \mathbf{\bar{H}}+ \frac{\mathrm{tr} \left\{  \mathbf{R}_n   \right\}}{E_T}  \right)^{-1}  \right\} = E_T
\end{equation}

\noindent the expression for the scaling factor 

\begin{equation}
\beta_{MMSE} =  \sqrt{ \frac{E_T}{ \mathrm{tr} \left\{ \mathbf{A} \left(  \mathbf{\bar{H}}^H \mathbf{\bar{H}}+ \frac{\mathrm{tr} \left\{  \mathbf{R}_n   \right\}}{E_T}  \right)^{-1}  \mathbf{\bar{H}}^H  \mathbf{R}_x  \mathbf{\bar{H}} \left(  \mathbf{\bar{H}}^H \mathbf{\bar{H}}+ \frac{\mathrm{tr} \left\{  \mathbf{R}_n   \right\}}{E_T}  \right)^{-1} \right\} }  }
\end{equation}

\noindent Therefore, the general linear constrained MMSE precoding is given by 

\begin{align}
\mathbf{F}_{MMSE} = &  \sqrt{ \frac{E_T}{ \mathrm{tr} \left\{ \mathbf{A} \left(  \mathbf{\bar{H}}^H \mathbf{\bar{H}}+ \frac{\mathrm{tr} \left\{  \mathbf{R}_n   \right\}}{E_T}  \right)^{-1}  \mathbf{\bar{H}}^H  \mathbf{R}_x  \mathbf{\bar{H}} \left(  \mathbf{\bar{H}}^H \mathbf{\bar{H}}+ \frac{\mathrm{tr} \left\{  \mathbf{R}_n   \right\}}{E_T}  \right)^{-1} \right\} }  } \\ \notag
& \times \left( \mathbf{\bar{H}}^H  \mathbf{\bar{H}}+ \frac{\mathrm{tr} \left\{    \mathbf{R}_n   \right\}}{E_T} \mathbf{A} \right)^{-1} \mathbf{\bar{H}}^H 
\end{align}


\section*{Acknowledgment}
The authors would like to thank Professor Lukas Landau from the Pontifical Catholic University of Rio de Janeiro for his contributions related to the derivation of the appropriate expression of the MMSE precoder for mmWave.

\IEEEtriggeratref{20}

\bibliographystyle{IEEEtran}
\bibliography{IEEEabrv,bibliography}

\begin{thebibliography}{10}
\providecommand{\url}[1]{#1}
\csname url@samestyle\endcsname
\providecommand{\newblock}{\relax}
\providecommand{\bibinfo}[2]{#2}
\providecommand{\BIBentrySTDinterwordspacing}{\spaceskip=0pt\relax}
\providecommand{\BIBentryALTinterwordstretchfactor}{4}
\providecommand{\BIBentryALTinterwordspacing}{\spaceskip=\fontdimen2\font plus
\BIBentryALTinterwordstretchfactor\fontdimen3\font minus
  \fontdimen4\font\relax}
\providecommand{\BIBforeignlanguage}[2]{{%
\expandafter\ifx\csname l@#1\endcsname\relax
\typeout{** WARNING: IEEEtran.bst: No hyphenation pattern has been}%
\typeout{** loaded for the language `#1'. Using the pattern for}%
\typeout{** the default language instead.}%
\else
\language=\csname l@#1\endcsname
\fi
#2}}
\providecommand{\BIBdecl}{\relax}
\BIBdecl

\bibitem{1}
O.~El~Ayach, S.~Rajagopal, S.~Abu-Surra, Z.~Pi, and R.~W. Heath, ``Spatially
  sparse precoding in millimeter wave {MIMO} systems,'' \emph{IEEE transactions
  on wireless communications}, vol.~13, no.~3, pp. 1499--1513, 2014.

\bibitem{8}
J.~Song, J.~Choi, and D.~J. Love, ``Common codebook millimeter wave beam
  design: Designing beams for both sounding and communication with uniform
  planar arrays,'' \emph{IEEE Transactions on Communications}, vol.~65, no.~4,
  pp. 1859--1872, 2017.

\bibitem{29}
C.~Hu, J.~Liu, X.~Liao, Y.~Liu, and J.~Wang, ``A novel equivalent baseband
  channel of hybrid beamforming in massive multiuser {MIMO} systems,''
  \emph{IEEE Commun. Lett., vol. PP}, no.~99, pp. 1--1, 2017.

\bibitem{3_12_11}
A.~Alkhateeb, O.~El~Ayach, G.~Leus, and R.~W. Heath, ``Hybrid precoding for
  millimeter wave cellular systems with partial channel knowledge,'' in
  \emph{Information Theory and Applications Workshop (ITA), 2013}.\hskip 1em
  plus 0.5em minus 0.4em\relax IEEE, 2013, pp. 1--5.

\bibitem{1just}
J.~Lee, G.-T. Gil, and Y.~H. Lee, ``Channel estimation via orthogonal matching
  pursuit for hybrid {MIMO} systems in millimeter wave communications,''
  \emph{IEEE Transactions on Communications}, vol.~64, no.~6, pp. 2370--2386,
  2016.

\bibitem{2just}
J.~Mirza, B.~Ali, S.~S. Naqvi, and S.~Saleem, ``Hybrid precoding via successive
  refinement for millimeter wave {MIMO} communication systems,'' \emph{IEEE
  Communications Letters}, vol.~21, no.~5, pp. 991--994, 2017.

\bibitem{3_14}
T.~E. Bogale and L.~B. Le, ``Beamforming for multiuser massive {MIMO} systems:
  Digital versus hybrid analog-digital,'' in \emph{Global Communications
  Conference (GLOBECOM), 2014 IEEE}.\hskip 1em plus 0.5em minus 0.4em\relax
  IEEE, 2014, pp. 4066--4071.

\bibitem{n11_8}
R.~M{\'e}ndez-Rial, C.~Rusu, N.~G. Prelcic, A.~Alkhateeb, and R.~W. Heath~Jr,
  ``Hybrid mimo architectures for millimeter wave communications: Phase
  shifters or switches?'' \emph{IEEE Access}, vol.~4, no.~8, pp. 247--267,
  2016.

\bibitem{n11_9}
M.~Kim and Y.~H. Lee, ``Mse-based hybrid rf/baseband processing for
  millimeter-wave communication systems in {MIMO} interference channels,''
  \emph{IEEE Transactions on Vehicular Technology}, vol.~64, no.~6, pp.
  2714--2720, 2015.

\bibitem{n11_10}
X.~Yu, J.-C. Shen, J.~Zhang, and K.~B. Letaief, ``Alternating minimization
  algorithms for hybrid precoding in millimeter wave {MIMO} systems.'' \emph{J.
  Sel. Topics Signal Processing}, vol.~10, no.~3, pp. 485--500, 2016.

\bibitem{21}
A.~Li and C.~Masouros, ``Hybrid precoding and combining design for
  millimeter-wave multi-user {MIMO} based on {SVD},'' in \emph{Communications
  (ICC), 2017 IEEE International Conference on}.\hskip 1em plus 0.5em minus
  0.4em\relax IEEE, 2017, pp. 1--6.

\bibitem{m3}
{\'A}.~J. Ortega, R.~Sampaio-Neto, and R.~P. David, ``Hybrid precoded index
  modulation in downlink mmwave {MU-MIMO} systems,'' in \emph{2019
  International Conference on Computing, Networking and Communications
  (ICNC)}.\hskip 1em plus 0.5em minus 0.4em\relax IEEE, 2019, pp. 329--333.

\bibitem{13}
G.~Kwon and H.~Park, ``A joint scheduling and millimeter wave hybrid
  beamforming system with partial side information,'' in \emph{Communications
  (ICC), 2016 IEEE International Conference on}.\hskip 1em plus 0.5em minus
  0.4em\relax IEEE, 2016, pp. 1--6.

\bibitem{29_8}
Z.~Pi, ``Optimal transmitter beamforming with per-antenna power constraints,''
  in \emph{Communications (ICC), 2012 IEEE International Conference on}.\hskip
  1em plus 0.5em minus 0.4em\relax IEEE, 2012, pp. 3779--3784.

\bibitem{BDdelamare}
K.~Zu, R.~C. de~Lamare, and M.~Haardt, ``Generalized design of low-complexity
  block diagonalization type precoding algorithms for multiuser {MIMO}
  systems,'' \emph{IEEE Transactions on Communications}, vol.~61, no.~10, pp.
  4232--4242, 2013.

\bibitem{n12}
M.~Joham, W.~Utschick, and J.~A. Nossek, ``Linear transmit processing in {MIMO}
  communications systems,'' \emph{IEEE Transactions on signal Processing},
  vol.~53, no.~8, pp. 2700--2712, 2005.

\bibitem{n11}
S.~S. Ioushua and Y.~C. Eldar, ``Hybrid analog-digital beamforming for massive
  {MIMO} systems,'' \emph{arXiv preprint arXiv:1712.03485}, 2017.

\bibitem{elements_theory}
T.~M. Cover and J.~A. Thomas, \emph{Elements of information theory}.\hskip 1em
  plus 0.5em minus 0.4em\relax John Wiley \& Sons, 2012.

\bibitem{n10}
A.~Scaglione, P.~Stoica, S.~Barbarossa, G.~B. Giannakis, and H.~Sampath,
  ``Optimal designs for space-time linear precoders and decoders,'' \emph{IEEE
  Transactions on Signal Processing}, vol.~50, no.~5, pp. 1051--1064, 2002.

\end{thebibliography}


\begin{IEEEbiography}
    [{\includegraphics[width=1in,height=1.25in,clip,keepaspectratio]{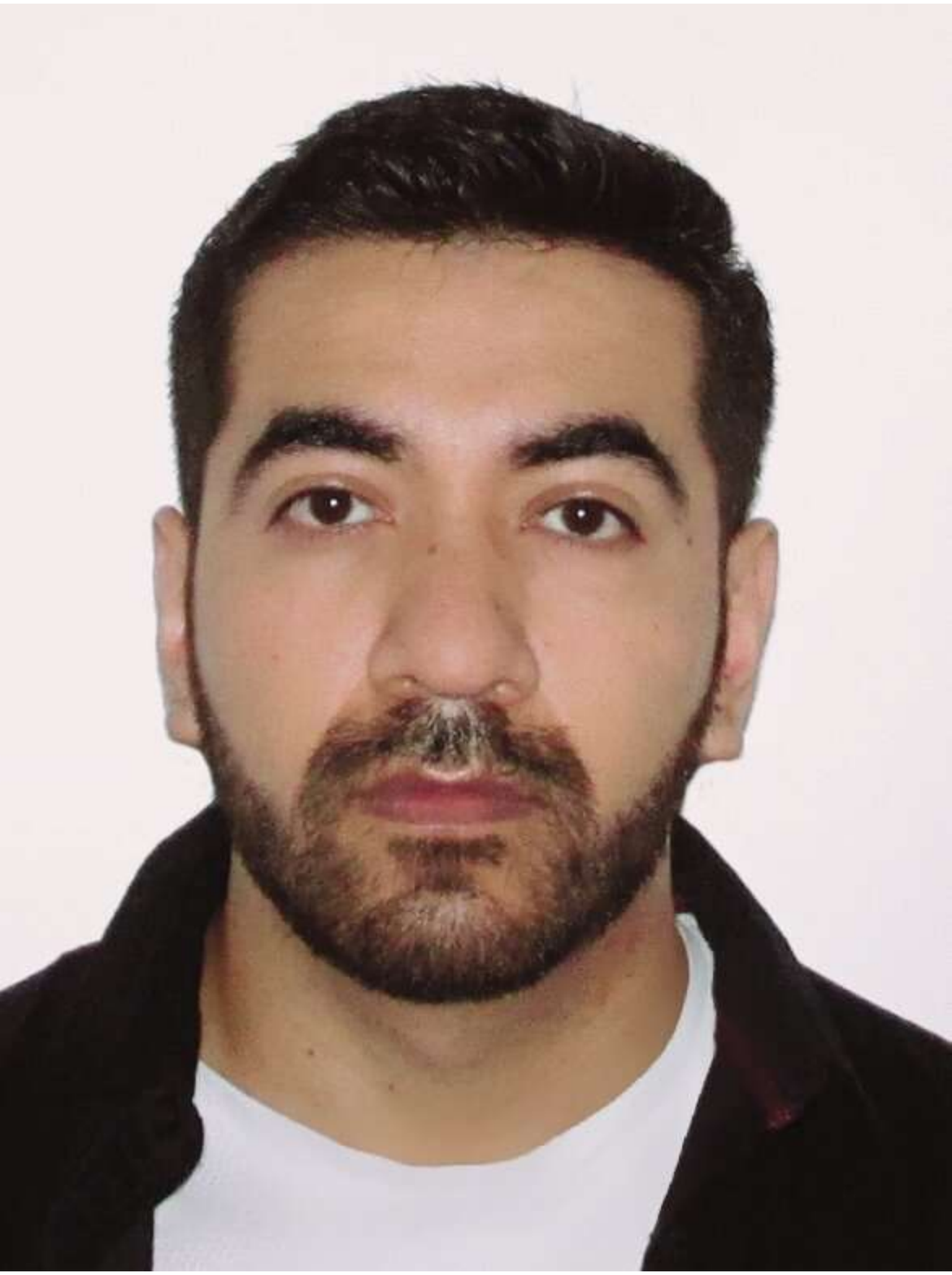}}]{Alvaro Javier Ortega}
graduated in Electronic Engineering in 2013 at the University of Nariño, Nariño, Colombia. In 2016, he obtained a Master's degree in Electrical Engineering Sciences from the Pontifical Catholic University of Rio de Janeiro (PUC-Rio), Rio de Janeiro, Brazil, where, due to his outstanding academic performance he was granted a special scholarship Grade 10, FAPERJ. From February 2016 to July 2016, he worked as Lecturer at the University of Nariño, Nariño, Colombia. Since August 2016, he is a Ph.D. student at the Electrical Engineering Department of PUC-Rio, Rio de Janeiro Brazil.  

His research interests include fifth-generation cellular networks, MIMO systems, hybrid processing, digital transmission and signal processing for communications.

\end{IEEEbiography}

\begin{IEEEbiography}
    [{\includegraphics[width=1in,height=1.25in,clip,keepaspectratio]{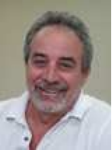}}]{Raimundo Sampaio-Neto}
received the Diploma and M.Sc. degrees from the Pontifical Catholic University
of Rio de Janeiro (PUC-Rio), Rio de Janeiro, Brazil, in 1975 and 1978, respectively, and the Ph.D. degree from the University of Southern California (USC), Los Angeles, CA, USA, in 1983, all in electrical engineering.
From November 1983 to June 1984, he was a Postdoctoral Fellow with the Communication Sciences
Institute, Department of Electrical Engineering, USC, and a Member of the Technical Staff with
Axiomatic Corporation, Los Angeles. He is currently a Researcher with the Center for Studies in Telecommunications and an Associate Professor with the Department of Electrical Engineering, PUC-Rio, where he has been since July 1984. In 1991, he was a Visiting Professor with the Department of Electrical Engineering, USC. He has participated in various projects and consulted for several private companies and government agencies. His areas of interest
include satellite communications, communication theory, digital transmission and signal processing for communications, areas in which he has published more than 180 papers in referred journals and conferences.
Dr. Sampaio-Neto was a co-organizer of the Session on Recent Results for the 1992 IEEE Workshop on Information Theory in Salvador, Brazil. He also served as a Technical Program Cochairman for the 1999 IEEE Global Telecommunications Conference held in Rio de Janeiro and as a Technical Program Committee Member for several national and international conferences. He was in office for three terms on the Board of Directors of the Brazilian Communications Society (SBrT). He served as a member of its Advisory Council for four terms and as an Associate Editor for the society journal: the \textit{Journal of Communication and Information Systems}. He is currently an Emerit Member of SBrT.

\end{IEEEbiography}


\begin{IEEEbiography}
    [{\includegraphics[width=1in,height=1.25in,clip,keepaspectratio]{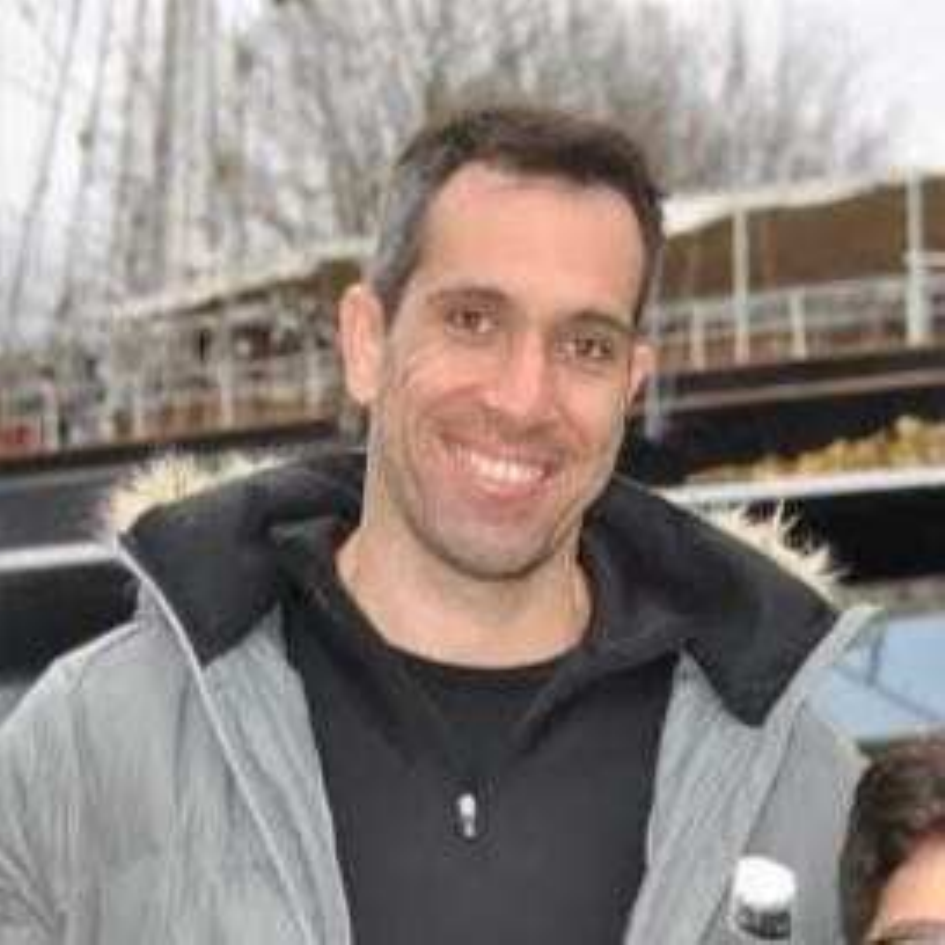}}]{Rodrigo Pereira David}
Graduate in Electrical Engineering from the State University of Rio de Janeiro (2002), MSc (2007) and PhD (2014) in Communication Systems from the Pontifical Catholic University of Rio de Janeiro. He is currently researcher at Laboratory of Telecommunications in the Brazilian Institute of metrology (INMETRO). Have experience in Electrical Engineering with emphasis on Telecommunication Systems, acting on the following topics: satellite communication systems, signal processing for communications, digital broadcasting and wireless communications systems.
\end{IEEEbiography}

\end{document}